\documentclass[10pt,journal,compsoc]{IEEEtran}

\ifCLASSOPTIONcompsoc
  \usepackage[nocompress]{cite}
\else
  \usepackage{cite}
\fi

\usepackage{amsmath}
\usepackage{graphicx}

\usepackage{amssymb}
\usepackage{subfigure} 
\usepackage{algorithm}
\usepackage{algpseudocode}

\usepackage{makecell}
\usepackage{balance}

\begin{document}

\title{Closed Walk Sampler: An Efficient Method for Estimating Eigenvalues of Large Graphs}


\author{Guyue~Han and Harish~Sethu
\IEEEcompsocitemizethanks{\IEEEcompsocthanksitem The authors are with the Department
of Electrical and Computer Engineering, Drexel University, Philadelphia, PA 19104.\protect\\
Email: \{guyue.han, sethu\}@drexel.edu}
}


%

\IEEEtitleabstractindextext{%
\begin{abstract}
Eigenvalues of a graph 
are of high interest in graph analytics for Big Data due to their relevance to many important properties of the graph including network resilience, community detection and the speed of viral propagation. Accurate computation of eigenvalues of
extremely large graphs is usually not feasible due to the prohibitive
computational and storage costs and also because full access to many
social network graphs is often restricted to most researchers. In this
paper, we present a series of new sampling algorithms which solve both of the above-mentioned problems and estimate the two largest eigenvalues of a large graph efficiently and with high accuracy. Unlike
previous methods which try to extract a subgraph with the most
influential nodes, our algorithms sample only a small portion of the
large graph via a simple random walk, and arrive at estimates of
the two largest eigenvalues by estimating the number of closed walks of a
certain length. Our experimental results using real graphs show that
our algorithms are substantially faster while also achieving significantly 
better accuracy on most graphs than the current state-of-the-art
algorithms.
\end{abstract}

}

\maketitle

\IEEEdisplaynontitleabstractindextext

%
\IEEEpeerreviewmaketitle

\IEEEraisesectionheading{\section{Introduction}\label{sec:introduction}}
\IEEEPARstart{S}{pectral} graph theory, which studies the spectral properties of the
Laplacian matrix or the adjacency matrix of a graph, plays an
important role in BigData analytics of large graphs
\cite{Chung1997}. Eigenvalues of a graph (the graph spectrum) can be
shown to be related to many principal properties of a graph and have
always had applications in chemistry, physics and other applied
sciences where graphs are studied and analyzed. In information theory,
the channel capacity can be defined in terms of the eigenvalues of the
channel graph \cite{Cohn1995}. In quantum chemistry, the graph
spectrum and the corresponding eigenvalues are highly relevant to the
stability of the molecule
\cite{Hong1993}. In Big Data applications involving graphs, such as
indexing for web search or social network analysis, eigenvalues of the
adjacency matrix can be helpful in characterizing graphs in a variety
of ways \cite{CveRow1997,BriPag1998,Chung2006}.  

In this paper, we focus on the largest and the second largest eigenvalues of the adjacency
matrix of the graph. These two eigenvalues have drawn much attention and have been
studied extensively for their relationship to multiple graph properties
of high relevance. The propagation properties of a graph can be
captured by the largest eigenvalue; as presented in \cite{WanCha2003},
an epidemic dies out when the curing rate is larger than the product
of the birth rate and the largest eigenvalue. The largest eigenvalue
is important to applications related to network robustness, community
detection and traffic engineering \cite{MahKri2006}. Mixing time, the number of steps that a random walk takes to arrive at stationary distribution, is related to the second largest eigenvalue \cite{Lovasz1993}. The spectral gap,
the difference between the largest and second largest eigenvalues, can
estimate the conductance of the network and describes the
connectivity, expansion and randomness properties of the graph
\cite{BroHae2011,MahKri2006}. 

While the study of the largest and the second largest eigenvalues has attracted much research,
efficient computation of these eigenvalues in case of massive graphs
remains an unsolved problem. The extremely large size of many graphs
of interest today (e.g., social network graphs) makes it difficult, or
sometimes even infeasible, to compute certain complex properties of
these graphs such as its eigenvalues. Power iteration, one of the most
famous and widely used algorithms for calculating the largest
eigenvalue and its associated eigenvector, requires $O(|E|)$ at each
iteration ($|E|$ is the number of edges in the graph)\cite{NobDan1988}. 
This method can also be used to calculate the second largest eigenvalue if the eigenvector of the largest eigenvalue is given or is calculated first.

Restricted access to the full graph is the other barrier to
researchers being able to compute the eigenvalues of large
graphs. The complete structural information of most social network
graphs (e.g., Facebook) is hidden except to privileged users with
access to the internal servers of the companies hosting the
network. Thankfully, however, on most online social networks, the
neighboring nodes of a given node can be queried via its API for
developers. This feature enables a random walk on the graph and
becomes one of the only means by which a large restricted-access graph
can be studied for its most interesting properties such as its eigenvalues.

The goal of this paper is to develop new sampling algorithms which
overcome the two obstacles mentioned above, the prohibitive
computational and storage costs and the matter of restricted access to
the entire graph. This work proposes new efficient algorithms which
estimate the two largest eigenvalues of a large graph by sampling only a
small fraction of the graph by means of a random walk.

\subsection{Contributions}

A closed walk or a closed path on a graph is a sequence of nodes
starting and ending at the same node. Our contribution exploits the
fact that the number of closed walks of length $k$ is
equal to the $k$-th spectral moment of a graph. Thus, estimating the number of closed
walks of length $k$ in a large graph allows us to estimate the top eigenvalues of the graph. Based on this principle, we present a series of new sampling algorithms with increasing generalizations. They carry the name {\em Closed Walk Sampler}, abbreviated as {\em cWalker}, and can estimate the top eigenvalues of a graph by visiting only a small fraction of the graph via a random walk.


Section \ref{sec:rationale} presents the theoretical foundation behind
the Closed Walk Sampler. We show that the largest eigenvalue of the 
graph can be inferred from the probability with which a closed path of
length $k$ is observed in the random walk. We examine the variance and
the confidence interval of our estimate of the number of closed paths
in order to illustrate the issue of large deviations in the estimate
when observations of a closed path become rare. This section builds
the rationale for increasing the probability of observing closed
walks in the random walk. 

In Section \ref{sec:est-with-detect}, we propose cWalker-A,
which accepts a parameter $k$ and uses an estimate of the number of
closed walks of length $k$ to return an estimate of the largest
eigenvalue of the graph. This version of our algorithm examines all
the neighbors of nodes visited during the random walk to see if it can
find a closed path without directly traversing a closed path in the
random walk, thus increasing the probability of observing closed
paths. The cWalker-A algorithm is named cWalker-limited in our preliminary work \cite{HanSet2017Closed}.

Section \ref{sec:alg-kfinding} presents cWalker-B, a generalization of cWalker-A. It is named the cWalker algorithm in our preliminary work \cite{HanSet2017Closed}. Instead of accepting a parameter $k$ as an input, it computes a reasonable value of $k$ which provides a good balance between the accuracy and the computational cost under the 
constraints of meeting a certain accuracy target. 
This section also describes the theoretical basis behind the algorithm. 

Section \ref{sec:general_app} presents a generalized approach to estimate the top $n$ eigenvalues of a graph iteratively. In Section \ref{sec:cwalker_toptwo}, we propose the cWalker-C algorithm, a generalization of cWalker-B based on eigenvalues. It estimates the two largest eigenvalues at the same time.

In Section \ref{sec:performance}, we present a performance analysis of our algorithms, cWalker-B and cWalker-C, against other state-of-the-art algorithms. 
Section \ref{sec:conclude} concludes the paper. 

\subsection{Related Work}

A large amount of work has focused on computing the eigenvalues and
their associated eigenvectors of matrices and graphs. The naive method
for exactly computing the eigenvalues of a matrix needs to find the
roots of the characteristic polynomial of the matrix
\cite{Chung1997}. However, computing the roots of the characteristic
polynomial of even a small matrix can be expensive and time-consuming,
which makes it computationally infeasible for the adjacency matrices
of large graphs. 

One class of approaches has tried to develop algorithms which produce
approximations to the eigenvalues and associated eigenvectors\cite{KucWoz1992,
  BatWil1976,PetWil1979,Parlet1974,Wilkin1965}. These algorithms are
iterative, with better approximations at each new iteration. The Power
Iteration is among the most famous and popular iterative algorithms
for finding the largest eigenvalue and its associated eigenvector
\cite{NobDan1988}. The iteration is terminated when two consecutively
calculated values of the largest eigenvalue are sufficiently close.  

Besides the Power Iteration, a number of other iterative algorithms
and their variations have been widely studied and have been used in
research. Subspace Iteration \cite{BatWil1973,BatRam1980} can produce
several of the largest eigenvalues and associated eigenvectors of a
symmetric matrix. Inverse Iteration \cite{PetWil1979,Ipsen1997} and
Rayleigh Quotient Iteration \cite{Parlet1974} are modifications of the
Power Iteration. They require fewer iterations, and obtain a faster
convergence than the original Power Iteration. QR
algorithm\cite{Wilkin1965} computes all eigenvalues and associated
eigenvectors; due to its complexity, it is often applied only to small
matrices. An overview of some popular iterative methods for computing
eigenvalues and eigenvectors, along with a summary of their advantages
and drawbacks, can be found in \cite{Panju2011}. 

All of these iterative methods require the complete information about
the graph, while in our case we assume the reality we face in the
analysis of large social network graphs --- that the access to the
full graph is restricted. Unfortunately, therefore, none of the above
approaches can serve as a feasible solution to the problem of
estimating the eigenvalues of a large graph accessible only through a
limited API made available to developers. 

A different approach to understanding the eigenvalues of a graph is
through examining properties of a graph and inferring mathematical
bounds on them \cite{HonShu2001,Stanley1987,Nikifo2006,DasKum2004,CveSim1995}. However, these
bounds can only serve as a rough guide and are not tight enough to
provide an accurate estimate of the top eigenvalues. 

A third and more feasible approach to estimating complex properties of
large graphs is through sampling. Sampling approaches have been widely
used in research on estimating simple but key properties of
graphs such as the degree distribution, the global clustering
coefficient, centrality metrics, and motif statistics\cite{HanSet2016,JhaSes2013,WanLui2015}. One approach to
graph sampling has been through extracting, via sampling, a small
representative subgraph from the large graph \cite{AhmDuf2014} and
projecting the properties of the subgraph on to the complete
graph. While these graph sampling methods have largely focused on
simple graph properties, much less is known about sampling a large graph efficiently to estimate more complex properties such as its spectrum or even just its largest eigenvalue.

The body of research that comes closest to our work tries to find the
most influential nodes via eigenvalue centrality approximation. They
work by collecting these nodes into a subgraph sample and one can then
compute the largest eigenvalue of this subgraph to estimate the
largest eigenvalue of the full graph based on interlacing results in
spectral graph theory which allow one to bound the eigenvalues of the
full graph using the eigenvalues of its subgraphs. In
\cite{MaiBer2010}, the authors present the Expansion Sampling
algorithm (XS) which is capable of capturing various centralities
(including eigenvalue centrality) of the nodes. In this method, a
subgraph sample is maintained where a neighboring node of the subgraph
is added into the subgraph based on the number of its neighbors that
are neither in the current subgraph nor are the neighboring nodes of
the current subgraph sample. Cho \emph{et al.} propose the BackLink
Count (BLC) algorithm which collects nodes that have most neighbors
into its sample subgraph \cite{ChoGar1998}. In \cite{ChuSet2015}, the
authors propose a greedy algorithm called Spectral Radius Estimator
(SRE), which samples nodes with the largest neighborhood volume and
adds them into its subgraph. The algorithm tries to extract out of the
full graph a subgraph with as large a spectral radius as possible. 

While these algorithms based on sampling the nodes with the largest eigenvalue
centrality in the graph offer some promise, they all need to compute
some metric or a score, that is hypothesized to correspond to
eigenvalue centrality, for each neighboring node of the current sample
subgraph in order to select the node with the highest score. This leads to high
computational complexity. The algorithms proposed in this paper
avoid the computational and space complexity associated with such
calculations and estimate the largest and the second largest eigenvalues via a simple random
walk.

\section{The Rationale}
\label{sec:rationale}
In this section, we present the theoretical foundation for the Closed
Walk Sampler (cWalker). We illustrate the problem of large deviation
in the estimates made of the number of closed walks of any given
length and explain why we focus on increasing the probability of observing a
closed path during the random walk on the graph. 

\subsection{Preliminaries and Notation}
Consider a connected, undirected simple graph $G = (V,E)$ with node
set $V$ and edge set $E$. Let $v \in V$ denote a node in $G$ and let
$N(v)$ denote the set of neighbors of node $v$. Let $d(v)$ denote the
degree of node $v$ and let $D=\sum_{v\in V}d(v)$ denote the sum of the
degrees of all the nodes in $G$.

Let $A$ be the $|V| \times |V|$  adjacency matrix of graph $G$. Since
$G$ is an undirected graph, $A$ is symmetric and its eigenvalues are
all real. Let $\lambda_1 \geq \lambda_2\geq ...\geq \lambda_{|V|} $
denote the real eigenvalues of $A$. $\lambda_1$ is the largest
eigenvalue of its adjacency matrix.  The goal of our paper is to
estimate the value of $\lambda_1$ by visiting only a small portion of
the large graph via a random walk. 

Consider a random walk on $G$, $(r_1,r_2,r_3,...)$, where $r_1$ is the
starting node and $r_i$ denotes the node visited in step $i$. Let $t$
denote the mixing time of graph $G$. Mixing time is the number of
steps that a random walk takes to reach steady state distribution
\cite{Lovasz1993}. The mixing time describes how fast a random walk
converges to its stationary distribution.

Let $P^{(i)}(v_j)$ be the probability of visiting node
$v_j$ in step $i$. The probability of drawing a given node from the
stationary distribution is independent of the initial node chosen to
begin the random walk. Thus, for $i \ge t$ (random walk reaches the
mixing time), we can drop $i$ from the notation and let $P(v_j)$
denote the probability of visiting node $v_j$ in the stationary
distribution. As shown in \cite{Lovasz1993}, 
\begin{equation*}
P(v_j) = \frac{d(v_j)}{D}
 \end{equation*} 

Let ${\bf R}^{(k)}$ denote the set of all possible sequences of $k+1$ nodes which
can be traversed in a random walk in $G$; it is the set of all walks
of length $k$ (allowing repeated nodes) in $G$. Let $X^{(k)} = (x_1, x_2,..., x_k, x_{k+1})$ denote a sequence of $k+1$ nodes such that $X^{(k)} \in {\bf R}^{(k)}$. $X^{(k)}$ is a closed walk if $x_1=x_{k+1}$. 

Let $P(X^{(k)})$ denote the
probability that the random walk steps through exactly the sequence of nodes $X^{(k)}$. Then, $P(X^{(k)})$ is given by
\begin{eqnarray}
\label{eqn:selectionProb}
\nonumber P(X^{(k)}) & = &
                                         \frac{d(x_1)}{D}\frac{1}{d(x_1)}\frac{1}{d(x_{2})}
                               \cdots \frac{1}{d(x_{k})}\\                                       
&=&\left\{
\begin{array}{ll}
  {\displaystyle \frac{1}{D} \prod_{i=2}^{k}\frac{1}{d(x_i)} },  &    \mbox{$k> 1$ }\\ 
  {\displaystyle \frac{1}{D}},  &    \mbox{otherwise.}
\end{array} 
\right. 
\end{eqnarray}

Now, define the function $\omega(X^{(k)})$ as follows to indicate if $X^{(k)}$ is closed: 
\begin{equation*}
\omega(X^{(k)})=\left\{
\begin{array}{ll}
  1   &    \mbox{if $x_1=x_{k+1}$}\\ 
  0   &    \mbox{otherwise.}
\end{array} 
\right. 
\end{equation*}

Given any sequence of nodes visited in the random walk, the
probability of visiting any particular sequence can be calculated
using Eqn. (\ref{eqn:selectionProb}). The value of function
$\omega(X^{(k)})$ for the sequence can be obtained by checking if the
first and last nodes in the sequence are the same. 

\subsection{Relationship to the number of closed walks}
\label{sec:est-without}

An interesting fact about graph spectra is that the trace of the
$k$-th order of the adjacency matrix of a graph equals its $k$-th 
spectral moment \cite{CveRow1997}: 
\begin{equation}
\label{eqn:eigenTrace}
\mathrm{tr}(A^k) = \sum_{i=1}^{|V|} \lambda_i^k,
\end{equation}
where $A$ is the adjacency matrix of the graph and $\mathrm{tr}(A^k)$
denotes the trace of the matrix $A^k$.  

The number of closed walks of length $k$ in $G$ is equal to the trace of matrix $A^k$ \cite{CveRow1997}.
Therefore, we have
\begin{equation*}
 \sum_{X^{(k)} \in {\bf R}^{(k)}} \omega(X^{(k)}) = \mathrm{tr}(A^k)
\end{equation*}
Applying Eqn. (\ref{eqn:eigenTrace}),
\begin{equation}
\label{eqn:eigenWalk}
 \sum_{X^{(k)} \in {\bf R}^{(k)}} \omega(X^{(k)}) = \sum_{i=1}^{|V|} \lambda_i^k
\end{equation}
For large values of $k$, $\lambda_1^k$ becomes the dominant term in the RHS of Eqn. (\ref{eqn:eigenWalk}). Thus, we can get
\begin{equation}
\label{eqn:eigenapprox}
 \sum_{X^{(k)} \in {\bf R}^{(k)}} \omega(X^{(k)}) \approx \lambda_1^k
\end{equation}
The above equation shows that we can arrive at an approximation of the
largest eigenvalue $\lambda_1$ if we know the number of closed walks
of length $k$ in $G$. 

For any walk of length $k$, $X^{(k)}$, using  Eqn. (\ref{eqn:selectionProb}) we define function $f(X^{(k)})$ as follows,
\begin{equation}
f(X^{(k)}) = \frac{1}{P(X^{(k)})D} 
 \label{f-walkProb}
\end{equation}

Let ${R_i}^{(k)}=(r_{i-k},r_{i-k+1},...,r_i)$ denote a walk of length $k$ obtained in the random walk during steps $i-k$ through $i$, where ${R_i}^{(k)} \in {\bf R}^{(k)}$. Taking the expected value of $\omega({R_i}^{(k)})f({R_i}^{(k)})$, we have
\begin{align}
\label{eqn:est-closedwalk}
&\nonumber \mathrm{E}[\omega({R_i}^{(k)})f({R_i}^{(k)})]\\
&~~\nonumber = \sum_{X^{(k)} \in {\bf R}^{(k)} } P(X^{(k)})  \mathrm{E}[\omega({R_i}^{(k)})f({R_i}^{(k)})|{R_i}^{(k)}=X^{(k)}]\\
&~~\nonumber = \sum_{X^{(k)} \in {\bf R}^{(k)} } P(X^{(k)})\omega(X^{(k)})f(X^{(k)})\\
&~~=\frac{1}{D} \sum_{X^{(k)} \in {\bf R}^{(k)} }\omega(X^{(k)})
\end{align}

Using Eqns. (\ref{eqn:selectionProb}) and (\ref{eqn:eigenapprox}), we get
\begin{equation}
\label{eqn:eigenapprox_exp1}
\lambda_1^k \approx \left\{
\begin{array}{ll}
  {\displaystyle D \cdot \mathrm{E}\left[\omega({R_i}^{(k)}) \prod_{j=1}^{k-1} d(r_{i-j})\right]},   &    \mbox{$k>1$}\\ 
  D  \cdot \mathrm{E}[\omega({R_i}^{(k)})],  &    \mbox{$k=1$}
\end{array} 
\right. 
\end{equation}

According to Eqn. (\ref{eqn:eigenapprox_exp1}), we can come up with a
simple algorithm for estimating $\lambda_1$ by random walk. At each
step, we check if the previous $k+1$ nodes form a closed path. By
checking for a closed path during the random walk, we estimate
the number of closed walks of length $k$ in $G$. Then, we can easily
reach an approximation of  $\lambda_1$.

\subsection{Estimate of $D$}

As presented in Eqn. (\ref{eqn:eigenapprox_exp1}), the value of $D$, the
sum of the degrees of all the nodes in $G$ is required in order to
compute $\lambda_1$.  Since we assume that the access to the full
graph is restricted, the real value of $D$ remains unknown. However,
we can generate an estimate of the value of $D$ via random walk.

Consider the expected value of $1/d(r_i)$ over the random walk, where
$r_i \in V$ is the node visited in step $i$: 
  \begin{eqnarray}
E\left[\frac{1}{d(r_i)}\right] &=& \sum_{v \in V} \frac{d(v)}{D}\frac{1}{d(v)} =\frac{|V|}{D}
\label{eqn:appx-D}
\end{eqnarray}

Eqn. (\ref{eqn:appx-D}) suggests a naive way of estimating the value
of $D$. D is equal to the ratio of the total number of nodes in the
full graph to the expected value of the degree of the nodes visited in
the random walk. In this paper, we focus on the estimate of the
largest eigenvalue, so we assume that the total number of nodes is
already known. In many social networks, e.g., Flickr, the total number
of nodes is known.

When the total number of nodes is not actually known, some approaches
that estimate it via a random walk have been presented in
\cite{HarKat2013,KatLib2011,HarRic2009}. These approaches can be
easily combined with our method. So, in the case that we do not know
the number of nodes in advance, we still can estimate it and proceed
with our algorithm.

\subsection{Large deviation}
\label{sec:large-devi}
Theoretically, according to Eqn. (\ref{eqn:eigenapprox_exp1}), the
approximation of $\lambda_1$ is closer to the actual value if a larger
value of $k$ is applied. How does the selection of $k$ affect the
accuracy? Since the approximation of $\lambda_1$ is obtained from the
estimate of $\sum_{X^{(k)} \in {\bf R}^{(k)} } \omega(X^{(k)})$, here we
analyze the performance of estimating the number of closed walks of
length $k$ as a reference. 

Using Eqn. (\ref{eqn:est-closedwalk}), we get the variance of the
estimate of
\[
{\displaystyle \sum_{X^{(k)} \in {\bf R}^{(k)} } \omega(X^{(k)})}
\]
as follows:
\begin{align}
\nonumber &\mathrm{Var}[D\omega({R_i}^{(k)})f({R_i}^{(k)})] \\
\nonumber &= \mathrm{E}[(D \omega({R_i}^{(k)})f({R_i}^{(k)}))^2]-\mathrm{E}[D \omega({R_i}^{(k)})f({R_i}^{(k)})]^2\\
&=\sum_{X^{(k)} \in {\bf R}^{(k)} } \omega(X^{(k)})\left(\frac{1}{P(X^{(k)})}-1 \right)
\label{eqn:var-est}
\end{align}

The above equation shows that the deviation of the estimate becomes
larger as a larger value of $k$ is used. Let $\tilde{{\bf R}}^{(k)} $
denote the estimate of the number of closed walks of length $k$ in
$G$. A 95\% confidence interval for the estimate $\tilde{{\bf
    R}}^{(k)}$ is as follows:
\begin{equation*}
\left(\tilde{{\bf R}}^{(k)} - 1.96 \frac{\sigma}{\sqrt{n}}~,~~\tilde{{\bf R}}^{(k)} +1.96 \frac{\sigma}{\sqrt{n}}\right),
\end{equation*}
where $\sigma^2$ is the variance of $\tilde{{\bf R}}^{(k)}$, and $n$ is the length of the random walk.

As the above expression shows, the size of the confidence interval is
determined by $\sigma$ and $n$. This suggests that, for a larger value
of $k$, we have to increase the length of the random walk in order to
reach a better accuracy on the estimation of the number of closed
walks of  length $k$. 

Eqn. (\ref{eqn:var-est}) shows that the probability of visiting a
closed walk of length $k$ significantly affects the deviation of the
estimate. In many network graphs, the ratio of the number of closed
walks of a certain length $k$ to the total number of walks of that
length is very low for large $k$. It makes the observation of a closed
walk of a large length become a rare event, and thus leads to a large
deviation of the estimate. In order to improve the probability with
which we observe a closed walk of a given length, we propose the
cWalker-A, a basic version of our algorithm which examines paths
beyond the ones traversed by the random walk itself.

\section{Algorithm given $k$ (cWalker-A)}
\label{sec:est-with-detect}

In this section, we present cWalker-A, which estimates the
largest eigenvalue of a graph through a random walk based on
estimating the number of closed walks of a given length, $k$. The cWalker-B
algorithm presented in the next section generalizes the cWalker-A to
find the most appropriate length of closed walks to observe and upon
which to base the estimate of the largest eigenvalue.

In the naive method suggested by Eqn. (\ref{eqn:eigenapprox_exp1}),
an observation of a closed path in the random walk is confirmed by
checking whether the first and the last nodes in the path are the
same. It works fine when the value of $k$ is not too large. However,
as a larger value of $k$ is applied, the large deviation problem
becomes severe. The key to the solution of this problem is to increase
the probability of visiting a closed walk of any given length. Based on this
intuition, our approach checks if a path is closed by examining
the neighboring nodes of the penultimate nodes in the potential path.

Define the function $\phi(X^{(k)})$ as follows to indicate if it is
possible that, given a path $X^{(k)}$ traversed in a random walk, the
next step in the walk will lead to a traversed path $X^{(k+1)}$ which
is a closed walk: 
\begin{equation*}
\phi(X^{(k)})=\left\{
\begin{array}{ll}
  1   &    \mbox{if $x_1 \in N(x_{k+1})$}\\ 
  0   &    \mbox{otherwise.}
\end{array} 
\right. 
\end{equation*}
Note that this means that we can observe a closed path $X^{(k+1)}$ even
if the random walk does not actually traverse exactly the sequence of
nodes in $X^{(k+1)}$. By keeping track of neighbors of nodes visited
during the random walk, this method increases the probability that
closed walks will be observed.

\newcommand{\algrule}[1][.2pt]{\par\vskip.3\baselineskip\hrule height #1\par\vskip.3\baselineskip}
\begin{algorithm}[!t]
\caption{cWalker-A}\label{alg:eigen_alg}
\begin{algorithmic}[1]
\Require{Graph $G = (V,E)$, size of the graph $n$, length of closed walk $k$, random walk length $m$.}
\Ensure{Largest eigenvalue $\lambda_1$}
\State{$c \leftarrow 0$}
\State{$D_\mathrm{est} \leftarrow 0$}
\State{Start and continue random walk until after the mixing time $t$,
  reaching node $r_{i-1}$ at step $i-1$. $(i-1=t)$}
\While{$i < m$}
\State{$r_i \leftarrow $ Random node in $N(r_{i-1})$}
\State{$D_\mathrm{est} \leftarrow D_\mathrm{est} +\frac{1}{d(r_{i-1})} $}
\If{$r_{i-k} \in N(r_{i-1})$}
\State{$c \leftarrow c + p(R_i^{(k-1)}) $} 
\EndIf
\State{$i \leftarrow i+1$}
\EndWhile 
\State{$D_\mathrm{est} \leftarrow n (m-t)/D_\mathrm{est}$}
\State{$\lambda_1 \leftarrow \left(\frac{c   D_\mathrm{est}}{m-t}\right)^{1/k}$}
\State{\textbf{return} $\lambda_1$}
\end{algorithmic}
\end{algorithm} 

Similar to Eqn. (\ref{eqn:est-closedwalk}), we can obtain the following expected value,
\begin{align}
&\nonumber \mathrm{E}[\phi({R_i}^{(k)})f({R_i}^{(k)})]\\
&~~\nonumber = \sum_{X^{(k)} \in {\bf R}^{(k)} } P(X^{(k)})  \mathrm{E}[\phi({R_i}^{(k)})f({R_i}^{(k)})|{R_i}^{(k)}=X^{(k)}]\\
&~~\nonumber = \sum_{X^{(k)} \in {\bf R}^{(k)} } P(X^{(k)})\phi(X^{(k)})f(X^{(k)})\\
&~~=\frac{1}{D} \sum_{X^{(k)} \in {\bf R}^{(k)} }\phi(X^{(k)})
\end{align}

Note that function $\phi(X^{(k)})$ checks the occurrence of the closed
walk of length $k+1$. Thus, using Eqn. (\ref{eqn:eigenapprox}), we have
\begin{equation}
\label{eqn:eigenapprox_exp2}
\lambda_1^k \approx \left\{
\begin{array}{ll}
  {\displaystyle D \cdot \mathrm{E}\left[\phi({R_i}^{(k-1)}) \prod_{j=1}^{k-2} d(r_{i-j})\right]},   &    \mbox{$k>2$}\\ 
  D  \cdot \mathrm{E}[\phi({R_i}^{(k-1)})],  &    \mbox{$k=2$}
\end{array} 
\right. 
\end{equation}

Eqn. (\ref{eqn:eigenapprox_exp2}) suggests a way to encounter closed
walks without necessarily traversing those paths in the random
walk. At each step, we check if one of the neighboring nodes of the
current node is identical to the node visited $(k-1)$ steps
earlier. If it is, a closed walk of length $k$ is observed. Since
the random walk needs to query the neighborhood information of the
current node to decide the node visiting in the next step, our new
algorithm does not require any additional information gathering during
its walk.

Algorithm \ref{alg:eigen_alg} presents the pseudo code of
cWalker-A for estimating the largest eigenvalue $\lambda_1$. We
use variable $c$ to record the estimate of the number of closed walks
of length $k$ and $D_\mathrm{est}$ to store the estimate of $D$, the
sum of the degrees of all the nodes in the graph. 

After necessary initializations (lines 1--3), we start examining the
closeness of the paths we visited and recording the estimate of the
total degrees in the graph (lines 4--11). For clarity, we define here
the function $p(R_i^{(l)})$ as follows:
\begin{equation}
p(R_i^{(l)})=\left\{
\begin{array}{ll}
  {\displaystyle \prod_{j=2}^{l}d(r_{i-j})},   &    \mbox{$l> 1$ }\\ 
    1, &    \mbox{$l=1$}
\end{array} 
\right. 
\end{equation}
Line 12 computes the final estimate of the total degrees in $G$. 
Lines 13--14 compute the largest eigenvalue using
Eqn. (\ref{eqn:eigenapprox_exp2}) and return it. 

\begin{algorithm}[!t]
\caption{cWalker-B}\label{alg:eigen_alg2}
\begin{algorithmic}[1]
\Require{Graph $G = (V,E)$, size of the graph $n$, maximum length of closed walk $K$, random walk length $m$, accuracy target $\beta$.}
\Ensure{Largest eigenvalue $\lambda_1$}
\State{$c[k] \leftarrow 0$, $1\leq k\leq K$}
\State{$\lambda_1[k] \leftarrow 0$, $1\leq k\leq K$}
\State{$\alpha[k] \leftarrow 0.99$, $1\leq k\leq K$}
\State{$D_\mathrm{est} \leftarrow 0$}
\State{Start and continue random walk until after the mixing time $t$,
  reaching node $r_{i-1}$ at step $i-1$. $(i-1=t)$}
\While{$i < m$}
\State{$r_i \leftarrow $ Random node in $N(r_{i-1})$}
\State{$D_\mathrm{est} \leftarrow D_\mathrm{est} +\frac{1}{d(r_{i-1})} $}
\For{$k$ in $[2,K]$}
\If{$r_{i-k} \in N(r_{i-1})$}
\State{$c[k] \leftarrow c[k]  +  p(R_i^{(k-1)})$}
\EndIf
\EndFor
\State{$i \leftarrow i+1$}
\EndWhile 
\State{$D_\mathrm{est} \leftarrow n (m-t)/D_\mathrm{est}$}
\For{$k$ in $[2,K]$}
\State{$\lambda_1[k] \leftarrow \left(\frac{c[k]   D_\mathrm{est}}{m-t}\right)^{1/k}$}
\EndFor
\For{$k$ in $[3,K]$}
\If{$\lambda_1[k]<\lambda_1[k-2]$}
\State{$\lambda_2 \leftarrow (\lambda_1[k-2]^{k-2}-\lambda_1[k]^{k-2})^{\frac{1}{k-2}}$}
\Else{ $\lambda_2 \leftarrow \lambda_1[k]$}
\EndIf
\State{$\alpha[k] \leftarrow \frac{\lambda_2}{\lambda_1[k]}$}
\EndFor
\State{$k^{\prime} \leftarrow \mathrm{min} \left(\mathrm{ceil}[\frac{\log(\beta)}{\log(\mathrm{min}(\alpha))}], K\right)$}
\State{$k^{\prime} \leftarrow \mathrm{max}(k^{\prime},5)$}
\State{\textbf{return} $\lambda_1[k^{\prime}]$}
\end{algorithmic}
\end{algorithm} 

\section{Algorithm using best $k$ (cWalker-B)}
\label{sec:alg-kfinding}
Section \ref{sec:est-with-detect} describes how to estimate the
largest eigenvalue for a given value, $k$, of the lengths of closed
walks. This section addresses the issue of choosing a suitable value
of $k$. In this section, we present cWalker-B, the more complete version of our algorithm, 
that can find a reasonable value of $k$ and estimate $\lambda_1$ based
on an estimation of the number of closed walks of length $k$.

Consider large values of $k$, where $\lambda_1^k$ and $\lambda_2^k$
become the dominant terms in the RHS of Eqn. (\ref{eqn:eigenTrace}). We have
\begin{equation*}
\mathrm{tr}(A^k) \approx \lambda_1^k + \lambda_2^k
\end{equation*}
Let $\alpha = \lambda_2 /\lambda_1$ denote the ratio of the second largest and the largest eigenvalue. Thus,
\begin{equation}
\mathrm{tr}(A^k) \approx (1+\alpha^k)\lambda_1^k
\end{equation}
The above equation shows that when $\alpha^k$ tends to $0$,
$\lambda_1$ is approximately equal to the $k$-th root of the total
number of closed walks of length $k$. Thus, in order to get a precise
approximation of $\lambda_1$, $\alpha^k$ should be as small as
possible. As $k$ increases, the value of $\alpha^k$
decreases. However, as discussed in Section \ref{sec:large-devi}, when 
using a very large $k$ in the algorithm, the accuracy of the estimate
may actually decrease because of the large deviation, requiring one to
increase the length of the random walk to achieve reasonable
accuracy. This presents us with a trade-off between the computational
cost and the accuracy. In the cWalker-B algorithm, we tackle this by
allowing an input into the algorithm that bounds the estimated
$\alpha^k$ by what we call an accuracy target, $\beta$, and we try to
find the smallest value of $k$ such that the estimated $\alpha^k$ is
lower than $\beta$.

So, $\beta$, the accuracy target, is the upper bound of
$\alpha^k$. Since $\alpha^k \leq \beta$, we can say:
\begin{eqnarray}
\label{eqn:optK-appox}
k \ge \log (\beta) / \log (\alpha)
\end{eqnarray}
The above inequality shows that $\lceil \log (\beta) / \log
(\alpha)\rceil$ is the smallest value of $k$ that makes the value of
$\alpha^k$ no greater than $\beta$, the given bound.  

Consider large values of $k$, where $\lambda_1^{(k-2)}$ and
$\lambda_2^{(k-2)}$ become the dominant terms, and $\lambda_1^{k}$
becomes the only dominant term in the RHS of
Eqn. (\ref{eqn:eigenTrace}). We have 
\begin{eqnarray*}
&\mathrm{tr}(A^{(k-2)}) \approx \lambda_1^{(k-2)} + \lambda_2^{(k-2)}\\
&\mathrm{tr}(A^k) \approx \lambda_1^k 
\end{eqnarray*}
Substituting $\lambda_1$ with $\mathrm{tr}(A^k)^{\frac{1}{k}}$,
\begin{eqnarray}
\label{eqn:lambda_2_appx}
{\lambda_2} \approx \left[\mathrm{tr}(A^{(k-2)}) - {\mathrm{tr}(A^k)}^{\frac{k-2}{k}}\right]^{\frac{1}{k-2}}
\end{eqnarray}
Using the above approximation, we can compute an approximate value of  $\lambda_2$, and thus obtain the value of $\alpha$. Having $\alpha$ and $\beta$, we can use Eqn. (\ref{eqn:optK-appox}) to compute the reasonable value of $k$ which provides a good balance between the accuracy and the computational cost.
 
Algorithm \ref{alg:eigen_alg2} presents the pseudo-code of the cWalker-B
algorithm for estimating the largest eigenvalue $\lambda_1$ using a
suitable value of $k$ given an accuracy target $\beta$. The main data
structures of the algorithm are described as follows: 
\begin{itemize}
\item Array $c[1...K]$: This is the array of counters. The element
  $c[i]$ in this array records the estimate of the number of closed
  walks of length $i$. 
\item Array $\lambda_1[1...K]$: The element $\lambda_1[i]$ in this
  array stores the approximation of $\lambda_1$ when the length of the
  walk used for checking if a path is closed is $i$.
\item Array $\alpha[1...K]$: The element $\alpha[i]$ in this array
  stores the estimate of the ratio of the second largest eigenvalue to
  the largest eigenvalue. 
\end{itemize}

Lines 1--5 perform necessary initializations. In lines 6--15, we start
estimating $\lambda_1$ for each value of $k$ in the given range and
collecting the data to also estimate $D$. 
Lines 16--19 compute an estimate of $D$ and the final estimate of
$\lambda_1$ for each value of $k$. Lines 20--26 compute $\alpha$,
the ratio of the second largest and the largest eigenvalue for each
$k$. Theoretically, with the increase in the value of $k$, the
estimate of $\alpha$ is decreased and is getting closer to the actual
value of $\alpha$. However, due to the large deviation and the limit
of the length of the random walk, the estimate of $\alpha$ starts
increasing when $k$ is larger than a certain value. 
So we select the minimum value of $\alpha$ as the correct
approximation, and use Eqn. (\ref{eqn:optK-appox}) to calculate
$k^{\prime}$, the reasonable value of $k$ under the accuracy target
$\beta$. 
In the pseudo code, the upper bound and the lower bound of the value
of $k^{\prime}$ are set. This guarantees the performance of our
algorithm in exceptional circumstances, such as when $\alpha$ tends to
$1$. Lines 27--29 calculate the value of $k$ which gives a good
estimate of $\lambda_1$ and return the corresponding $\lambda_1$. 

According to Eqn. (\ref{eqn:optK-appox}), the smallest value of $k$ is
determined by $\alpha$. When $\alpha$ is close to $1$, the value of
$k$ has to be very large in order to have an accurate estimate of
$\lambda_1$. As discussed in Section \ref{sec:large-devi}, with a
larger value of $k$, the length of the random walk has to be
increased. In other words, the rate of convergence of our algorithm is
determined by $\alpha$, the ratio of the second largest and the
largest eigenvalues of the graph. If $\alpha$ is very close to $1$,
our algorithm has to perform a longer random walk to reach an accurate
estimate. Almost all real graphs have an $\alpha$ substantially lower
than 1, but it is possible for a real graph to have an $\alpha$ close
to 1.

\section{A generalized approach}
\label{sec:general_app}
Sections \ref{sec:rationale}--\ref{sec:alg-kfinding} describe the theoretical foundation behind our approach and present the cWalker-A and cWalker-B algorithms for estimating the largest eigenvalue of a graph. In this section, we present a generalized approach which can estimate the top $n$ eigenvalues of a graph iteratively.

For large values of $k$, $\sum_{i=1}^{c} {\lambda_i}^k$ becomes the dominant term in the RHS of Eqn. (\ref{eqn:eigenTrace}). Thus, We have
\begin{equation*}
\mathrm{tr}(A^k) \approx {\lambda_1}^k + {\lambda_2}^k + ... + {\lambda_c}^k
\end{equation*}
Let $\alpha_{c} = \lambda_c / \lambda_{c-1}$ denote the ratio of the $c$-th largest and the $(c-1)$-th largest eigenvalue. Thus,
\begin{eqnarray}
\label{equ:gen1}
\mathrm{tr}(A^k) -\sum_{i=1}^{c-2}{\lambda_i}^k \approx (1+\alpha_{c}^k){\lambda_{c-1}}^k
\end{eqnarray}
 $\lambda_{c-1}$ is approximately equal to the $k$-th root of the LHS of Eqn. (\ref{equ:gen1}) when ${\alpha_c}^k$ tends to 0.

Consider large values of $k$, where we can get the following equations,
\begin{gather*}
\mathrm{tr}(A^{k-2}) \approx \sum_{i=1}^{c}{\lambda_i}^{k-2} \\
\mathrm{tr}(A^k)  \approx \sum_{i=1}^{c-1}{\lambda_i}^k
\end{gather*}
We can easily have
\begin{gather}
\lambda_{c-1} \approx \left[\mathrm{tr}(A^{k})-\sum_{i=1}^{c-2}{\lambda_i}^{k}\right]^{\frac{1}{k}} \label{eqn:approx_general1}\\
\lambda_c  \approx \left[ \mathrm{tr}(A^{k-2}) - \mathrm{tr}(A^{k})^{\frac{k-2}{k}}\right]^{\frac{1}{k-2}}
\label{eqn:approx_general2}
\end{gather}
Suppose the values of the first $c-2$ largest eigenvalues are known, we can have an approximate value of $\lambda_{c-1}$ using Eqn. (\ref{eqn:approx_general1}). Combining Eqns. (\ref{eqn:approx_general1}) and (\ref{eqn:approx_general2}), we can compute an approximate value of $\lambda_c$, and thus obtain the value of $\alpha_c$. Then, similar to the approach described in Section \ref{sec:alg-kfinding}, we can come up with a reasonable value of $k$ for estimating $\lambda_{c-1}$. The estimates of the first $c-2$ largest eigenvalues can be obtained by using the above method iteratively. Thus, we have a generalized approach for estimating the top $n$ eigenvalues in the graph. To achieve an estimate of the $c$-th largest eigenvalue, estimates of the first $c-1$ largest eigenvalues are used in the approximation, so the error is propagated. In other words, the estimate obtained by this approach becomes less accurate for eigenvalues which rank behind. 

\begin{algorithm}[!t]
\caption{cWalker-C}\label{alg:cWalker-TopTwo}
\begin{algorithmic}[1]
\Require{Graph $G = (V,E)$, size of the graph $n$, maximum length of closed walk $K$, random walk length $m$, accuracy target $\beta$.}
\Ensure{Two largest eigenvalues $\lambda_1$ and $\lambda_2$}
\State{$c[k] \leftarrow 0$, $1\leq k\leq K$}
\State{$\lambda_1[k] \leftarrow 0$, $1\leq k\leq K$}
\State{$\alpha[k] \leftarrow 0.99$, $1\leq k\leq K$}
\State{$D_\mathrm{est} \leftarrow 0$}
\State{Start and continue random walk until after the mixing time $t$,
  reaching node $r_{i-1}$ at step $i-1$. $(i-1=t)$}
\While{$i < m$}
\State{$r_i \leftarrow $ Random node in $N(r_{i-1})$}
\State{$D_\mathrm{est} \leftarrow D_\mathrm{est} +\frac{1}{d(r_{i-1})} $}
\For{$k$ in $[3,K]$}
\State{$c[k] \leftarrow c[k] + p(R_i^{(k-2)}) |N(r_{i-k+1}) \cap N(r_{i-1})|$}
\EndFor
\State{$i \leftarrow i+1$}
\EndWhile 
\State{$D_\mathrm{est} \leftarrow n (m-t)/D_\mathrm{est}$}
\For{$k$ in $[3,K]$}
\State{$\lambda_1[k] \leftarrow \left(\frac{c[k]   D_\mathrm{est}}{m-t}\right)^{1/k}$}
\EndFor
\For{$k$ in $[4,K]$}
\If{$\lambda_1[k]<\lambda_1[k-2]$}
\State{$\lambda_2 \leftarrow (\lambda_1[k-2]^{k-2}-\lambda_1[k]^{k-2})^{\frac{1}{k-2}}$}
\Else{ $\lambda_2 \leftarrow \lambda_1[k]$}
\EndIf
\State{$\alpha[k] \leftarrow \frac{\lambda_2}{\lambda_1[k]}$}
\EndFor
\State{$k^{\prime} \leftarrow \mathrm{min} \left(\mathrm{ceil}[\frac{\log(\beta)}{\log(\mathrm{min}(\alpha))}], K\right)$}
\State{$k^{\prime} \leftarrow \mathrm{max}(k^{\prime},5)$}
\If{$\lambda_1[k'-2] \geq \lambda_1[k']$}
\State{$\lambda_2 \leftarrow (\lambda_1[k'-2]^{k'-2} - \lambda_1[k']^{k'-2})^{\frac{1}{k'-2}}$}
\Else{ $\lambda_2 \leftarrow \lambda_1[k'] $}
\EndIf
\State{\textbf{return} $\lambda_1[k^{\prime}]$ and $\lambda_2$}
\end{algorithmic}
\end{algorithm} 

\section{Estimating two largest eigenvalues}
\label{sec:cwalker_toptwo}
The cWalker-A algorithm presented in Section \ref{sec:est-with-detect} provides a way to increase the probability of observing closed walks by checking if one of the neighboring nodes of the current node is identical to the node visited $(k-1)$ steps earlier. In this section, we improve this method by further increasing the probability of encountering closed walks of given lengths and present cWalker-C, the algorithm that can estimate the two largest eigenvalues at the same time.

\subsection{Increasing encounters of closed paths}
\label{sec:approach-limited2}
Define the function $\phi(X^{(k)})$ as follows to indicate the number of possible closed paths of length $k+2$ in which the given path $X^{(k)}$ is involved, where $X^{(k)}$ is in the middle of these closed paths (etc., the first node in $X^{(k)}$ is the second node in the potential path),
\begin{equation*}
\phi(X^{(k)})=|N(x_{1}) \cap N(x_{k+1})|
\end{equation*}

The above function suggests that we can observe multiple closed paths of length $k+2$ by checking the number of common neighbors between the first and the last node in a given path $X^{(k)}$. 

Similar to the derivation of Eqn. (\ref{eqn:eigenapprox_exp2}), we can have
\begin{equation*}
\lambda_1^k \approx \left\{
\begin{array}{ll}
  {\displaystyle D \cdot \mathrm{E}\left[\phi({R_i}^{(k-2)}) \prod_{j=1}^{k-3} d(r_{i-j})\right]},   &    \mbox{$k>3$}\\ 
  D  \cdot \mathrm{E}[\phi({R_i}^{(k-2)})],  &    \mbox{$k=3$}
\end{array} 
\right. 
\end{equation*}

The above equation suggests a way to further increase the probability of observing closed paths in a random walk. At each step, we check the number of common nodes between the neighborhood of the current node and the node visited $(k-2)$ steps earlier. The number of common nodes indicates the number of closed walks of length $k$ being observed. However, this approach needs to find common nodes in two sets, and this leads to higher computational complexity.


%

\subsection{The algorithm (cWalker-C)}
Algorithm \ref{alg:cWalker-TopTwo} presents the pseudo-code of the cWalker-C algorithm for estimating the two largest eigenvalues $\lambda_1$ and $\lambda_2$. Since the accuracy of the estimates of $\lambda_1$ affects the accuracy of the estimates of $\lambda_2$, in the task of estimating the two largest eigenvalues at the same time, we choose to use the approach proposed in the above subsection (Section \ref{sec:approach-limited2}) to encounter closed paths. It takes more computational cost but achieves higher accuracy. 

Similar to the cWalker-B algorithm, lines 1--5 perform necessary initializations, and lines 6--17 estimate $D$ and $\lambda_1$ for each value of $k$. Lines 18--24 compute $\alpha$, the ratio of the second largest and the largest eigenvalue for each
$k$. Line 25 calculates $k^{\prime}$, the value of $k$ used for estimating $\lambda_1$ under the given accuracy target $\beta$. Eqn. (\ref{eqn:lambda_2_appx}) provides a way to get an approximate value of $\lambda_2$ using the number of closed walks of length $k$ and $k-2$. Thus, in lines 27--30, we use this equation to compute $\lambda_2$. Since we choose $k'$ to estimate $\lambda_1$, the value of $k$ for estimating $\lambda_2$ must be no larger than $k'$. Besides, as we discussed before, the estimate is more accurate when using a larger $k$. So we choose $k'$, the largest value of $k$ which can be used, to compute $\lambda_2$. Line 31 returns the estimates of the two largest eigenvalues $\lambda_1$ and $\lambda_2$.

\section{Performance Analysis}
 \label{sec:performance}
 
\begin{table}[!b]
\newcommand{\tabincell}[2]{\begin{tabular}{@{}#1@{}}#2\end{tabular}}
\centering
\caption{Graph datasets used in the analysis.}
\vspace{0.05in}
\label{tab:graphdata}
\begin{tabular}{c|c|c|c|c|c}  \Xhline{1pt}
\tabincell{c}{Graph\\(LCC)}& \tabincell{c}{Nodes\\$|V|$} &\tabincell{c}{Edges\\$|E|$}&$\lambda_1$&$\lambda_2$  & $\lambda_3$ \\   \Xhline{1pt}
email-EuAll& 224,832& 339,925 & 102.54 & 87.39 &79.60\\ \hline
com-Youtube&1,134,890&2,987,624&210.40&169.43 &154.82\\ \hline
loc-gowalla & 196,591&950,327 &170.94 &110.96 &104.85\\  \hline
com-Amazon & 334,863&925,872&23.98&23.91 &23.28\\
\hline\end{tabular}
\end{table}
 
\begin{table}[!b]
\centering
\caption{The relative errors in the estimates of the largest eigenvalue.}
\vspace{0.05in}
\label{tab:rel_error}
\begin{tabular}{c|ccc} 
\Xhline{1pt}
  &\multicolumn{3}{c}{Largest eigenvalue $\lambda_1$}   \\ 
Graph   &\multicolumn{3}{c}{Relative error (\%)}\\ 
 \cline{2-4}
name &  cWalker-B &BLC & SRE  \\ \hline
email-EuALL&1.25& 42.68& 43.76 \\ \hline
com-Youtube&9.32&57.54 & 48.56\\ \hline
loc-gowalla&  7.47 &43.03& 36.95 \\ \hline
com-Amazon& 4.29& 1.55 & 0.03\\ 
\Xhline{1pt}
\end{tabular}
\end{table}

\begin{figure*}[!t]
\centering
     
       \includegraphics[width=2in]{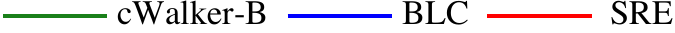}
       \\
    \subfigure[{email-EuAll}]{ 
       \includegraphics[width=2.85in]{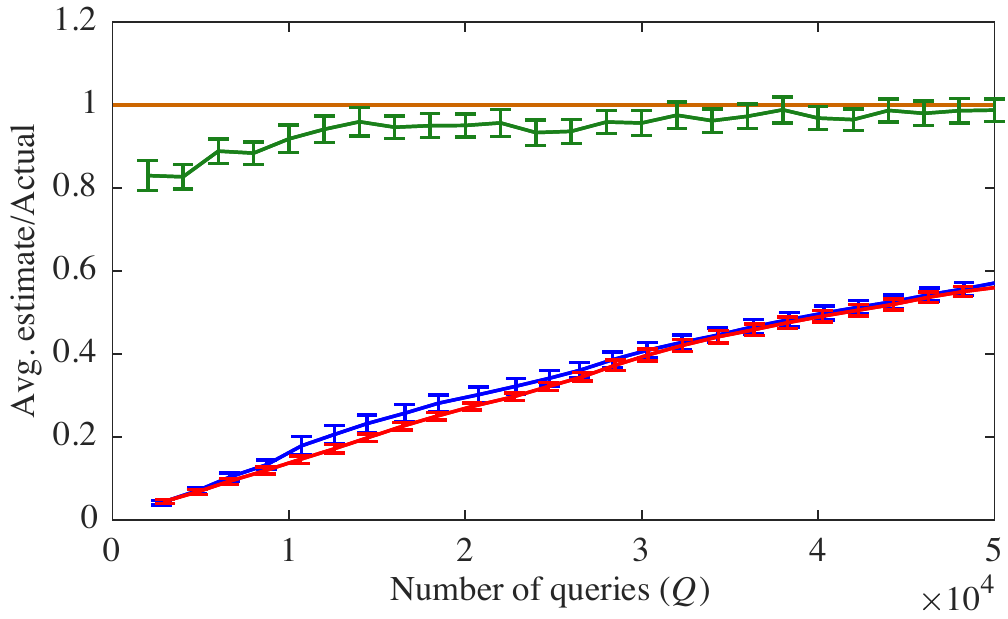} 
       }
      \subfigure[{com-Youtube}]{ 
       \includegraphics[width=2.85in]{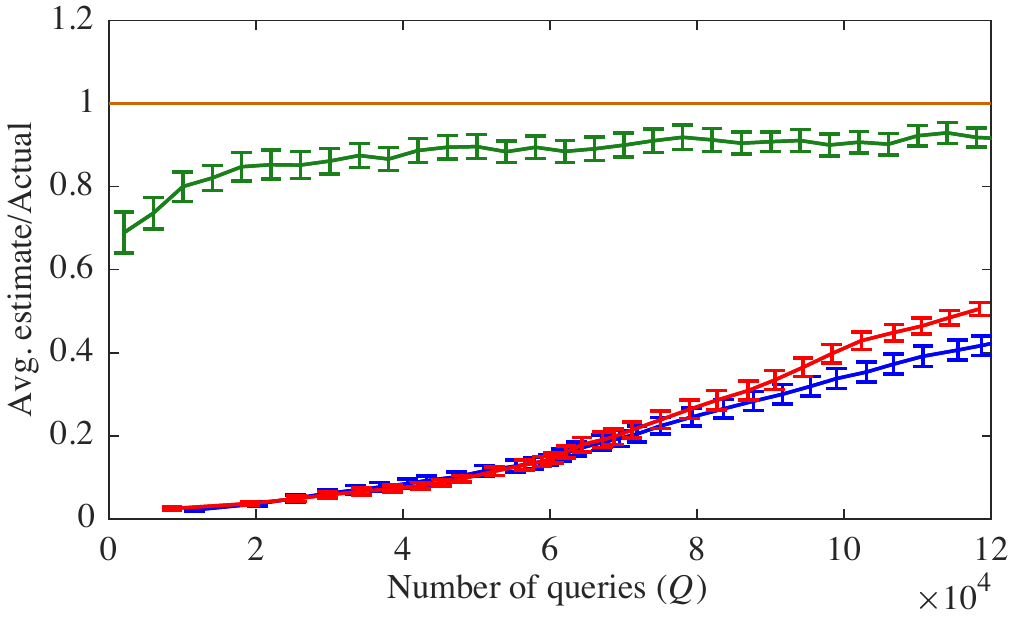}
       }
       \subfigure[{loc-gowalla}]{ 
       \includegraphics[width=2.85in]{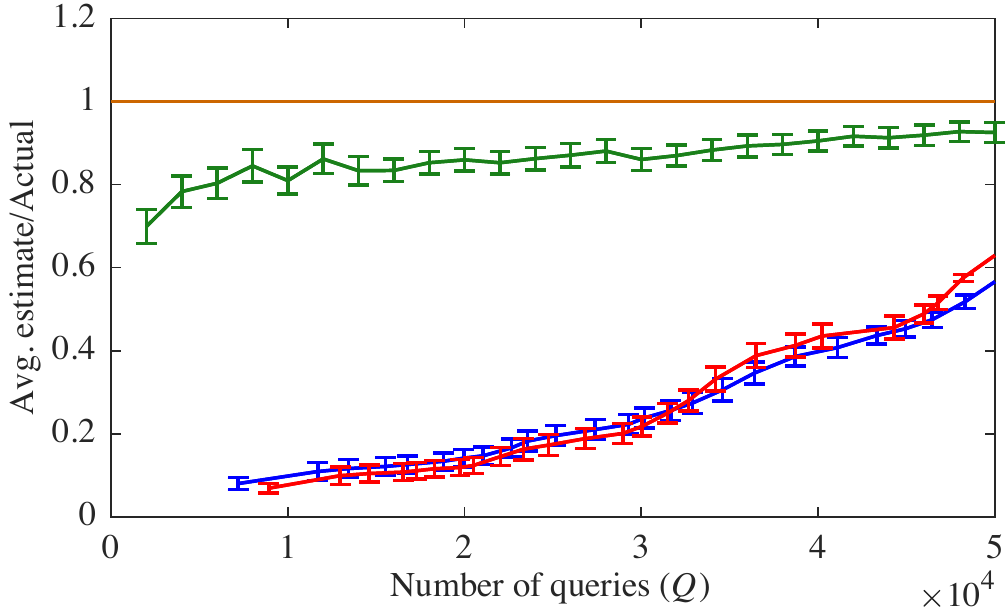}
       }
              \subfigure[{com-Amazon}]{ 
      \includegraphics[width=2.85in]{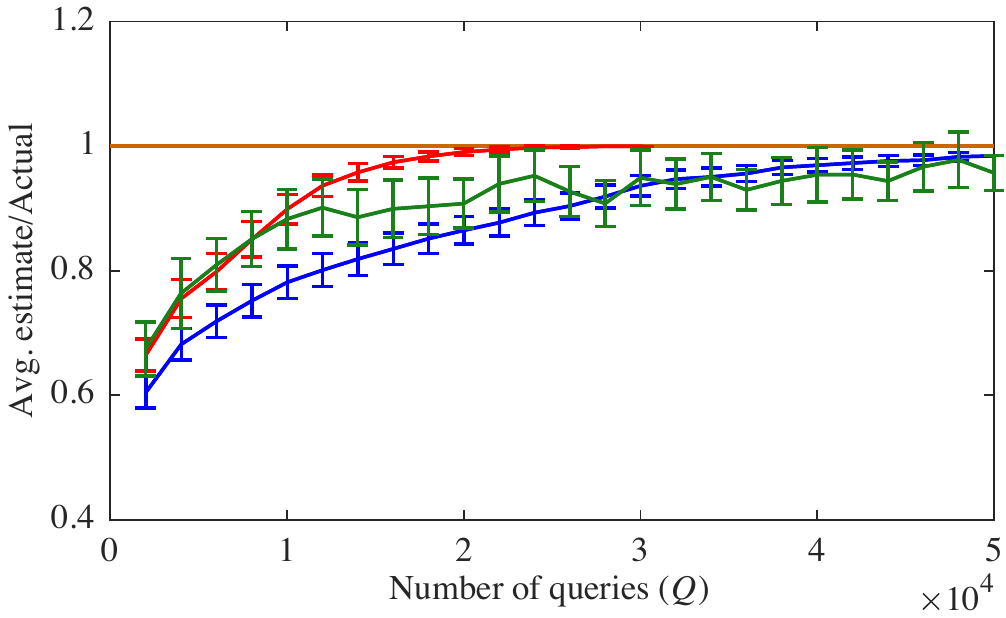}
      \label{fig:amazon-error}
        }

    \caption{Comparison of the ratio of the average estimated value of
      $\lambda_1$ and the actual value. The brown line indicates 1. The error bars indicate 95\% confidence intervals over 100 independent runs.}\label{fig:accuracy} 
\end{figure*} 

In this section, we present a performance analysis of cWalker-B and cWalker-C as described in Algorithm \ref{alg:eigen_alg2} and Algorithm  \ref{alg:cWalker-TopTwo}. We compare our algorithms 
against two state-of-the-art algorithms, Spectral Radius Estimator
(SRE) \cite{ChuSet2015} and BackLink Count (BLC)
\cite{ChoGar1998}. We do not consider XS algorithm \cite{MaiBer2010}
in this analysis because, as already established in \cite{ChuSet2015},
it performs substantially poorer than both SRE and BLC. 
Both SRE and BLC aim to find a set of nodes which have the largest estimated eigenvalue centrality. They estimate the largest eigenvalue of the original graph by calculating the largest eigenvalue of the subgraph induced by the set of sampled nodes with high eigenvalue centrality.

Our experiments were performed on real graphs from the Stanford
Network Analysis Project (SNAP) \cite{snapnets}. Table
\ref{tab:graphdata} lists some vital properties of these graphs. For
each graph used, the algorithms were run on the largest connected
component of the graph. 

\subsection{Results of estimating the largest eigenvalue}
In this subsection, we show results of estimating the largest eigenvalue. We compare the cWalker-B algorithm (as described in Algorithm \ref{alg:eigen_alg2}) against SRE and BLC. For all of the experiments of the cWalker-B algorithm, the accuracy target $\beta$ and the maximum length of closed walk $K$
were set to 0.05 and 30, respectively. 

\begin{figure*}[!t]
\centering
       \includegraphics[width=2in]{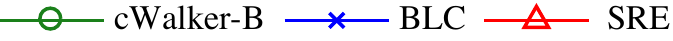}
       \\
    \subfigure[{email-EuAll}]{ 
       \includegraphics[width=2.85in]{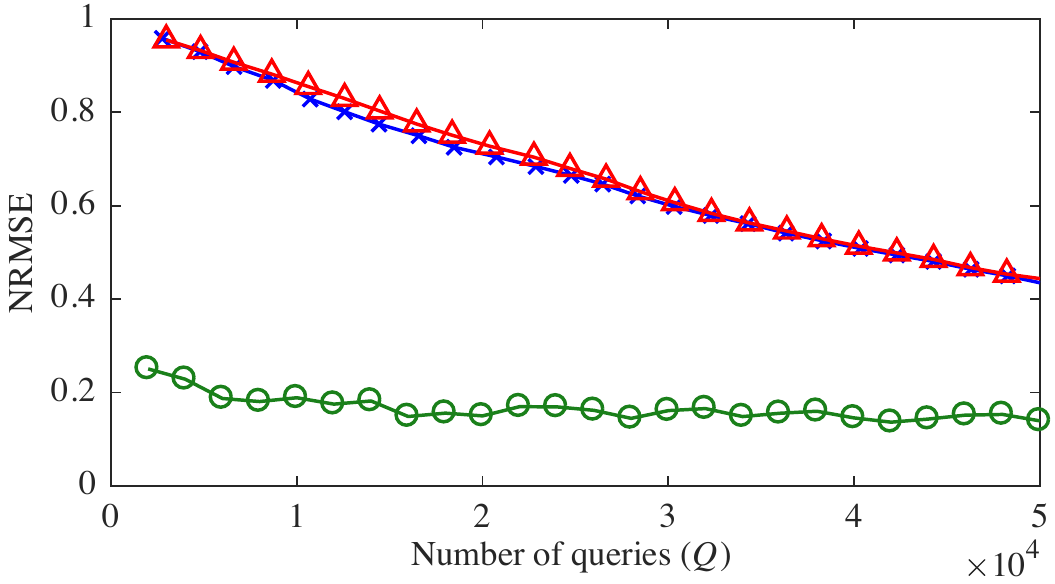}
       }
      \subfigure[{com-Youtube}]{ 
       \includegraphics[width=2.85in]{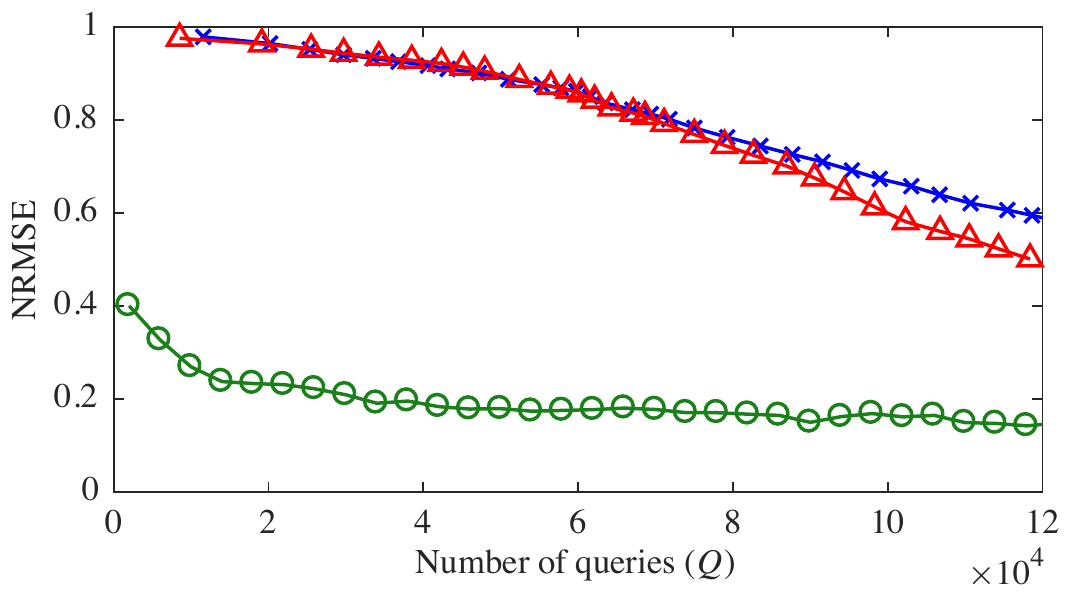}
       }
       \subfigure[{loc-gowalla}]{ 
       \includegraphics[width=2.85in]{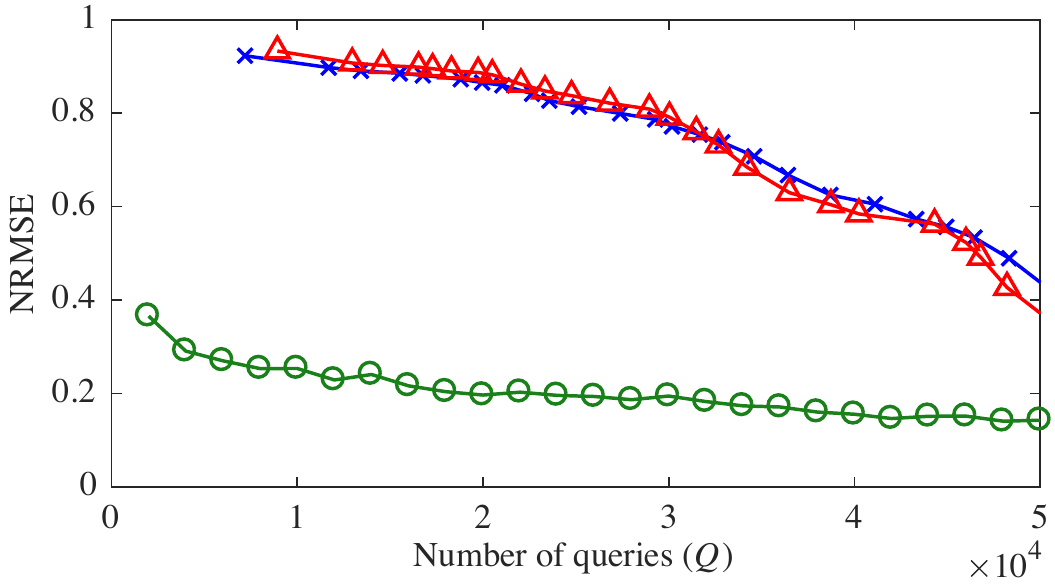}
       }
              \subfigure[{com-Amazon}]{ 
      \includegraphics[width=2.85in]{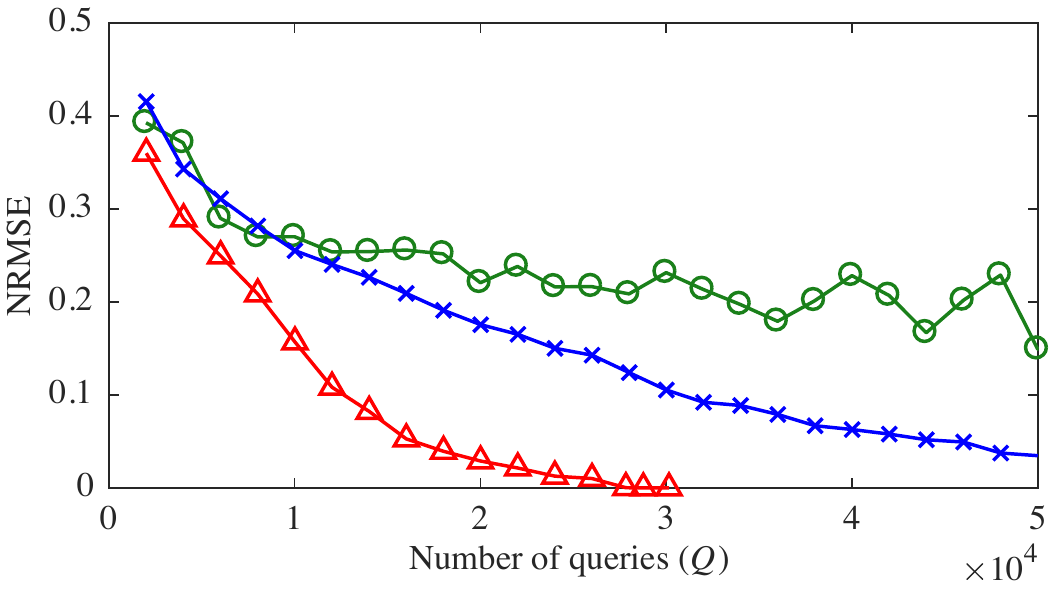}
      \label{fig:amazon-error}
        }

   \caption{Comparison of NRMSEs of the estimates by the three algorithms. }\label{fig:accuracy_nrmse} 
\end{figure*}

As described in Section \ref{sec:alg-kfinding}, the performance of our
algorithm is affected by $\alpha$, the ratio of the second largest and
the largest eigenvalue. The smaller the $\alpha$, the less information
it needs to converge to an answer with acceptable accuracy. For the
sake of completeness in our performance analysis, we demonstrate the
rare case when $\alpha$ is extremely close to 1 by deliberately
including the com-Amazon graph. Note that it is not common for real
graphs to have an $\alpha$ extremely close to 1. In fact, in our
study of 50 graphs listed on the SNAP site \cite{snapnets}, com-Amazon
graph had the highest value of $\alpha$ at 0.997. The $\alpha$ value
of the other graphs, we found, were between 0.38 and 0.98, with a mean
of 0.78 and a median of 0.82.
 
\subsubsection{\textbf{Accuracy}}
We consider the relative error as a measure of the accuracy. We
measure the relative error as: 
\[
\mathrm{Relative~error} = \frac{\mathrm{Average~estimate} -
  \mathrm{Actual~value}}{\mathrm{Actual~value}},
\]
where the average estimate is the mean of the estimated value across
100 independent runs. For each graph, we fixed $Q$, the number of
queries, to ensure that all of the three algorithms obtain the same
amount of information through its queries of the graph and to make
sure that the evaluation is under equivalent complexity. The SRE
algorithm is a greedy algorithm which keeps replacing the sampled
nodes with more influential nodes after the size of the sample graph
reaches the desired sample size. Here we set the desired size as 4\%
of the size of the original graph based on results in
\cite{ChuSet2015} which demonstrated a high accuracy at a sample size
set to 4\% of the full graph.

Table \ref{tab:rel_error} shows the relative errors in estimating the
largest eigenvalue for each of the three algorithms. The number of
queries is $120$K for the com-Youtube graph and $50$K for the other
graphs. As shown in the table, except for the case of the com-Amazon
graph, the relative errors achieved by our algorithm are significantly
better than BLC and SRE. 

A further comparison of the three algorithms is shown in
Fig. \ref{fig:accuracy}. It shows the ratio of the average estimated
$\lambda_1$ to the actual value for each of the four graphs as the
number of queries increased. As depicted in the figure, all of the
algorithms gradually converge to the actual value. In the cases of the
email-EuAll, com-Youtube and loc-gowalla graphs, our algorithm always
achieves substantially better accuracy than BLC and SRE with
increasing number of queries. 

To provide a more comprehensive analysis of the accuracy of the three
algorithms, we use the normalized root mean square error (NRMSE) which
infers both the variance and the bias of the estimates. The NRMSE
is defined as follows: 
\begin{eqnarray*}
\mathrm{NRMSE}=\frac{\sqrt{\mathrm{E}[(\mathrm{estimate}-\mathrm{Actual~value})^2]}}{\mathrm{Actual~value}},
\end{eqnarray*}
Fig. \ref{fig:accuracy_nrmse} plots the NRMSEs based on 100
independent runs for each graph with increasing number of
queries. Similar to the results plotted in Fig. \ref{fig:accuracy},
our algorithm performs significantly better in terms of accuracy in
the email-EuAll, com-Youtube and loc-gowalla graphs. 

As expected, the com-Amazon graph is the only case for which our algorithm
does not perform very well. We chose this graph as an exceptional
case to show the influence of $\alpha$ on the performance of our
algorithm. The ratio of $\lambda_2$ and $\lambda_1$, of the com-Amazon
graph is extremely close to $1$. As listed in Table
\ref{tab:graphdata}, the largest and the second largest eigenvalue of
the com-Amazon graph are 23.98 and 23.91, respectively.
Thus, as plotted in Fig. \ref{fig:amazon-error}, our algorithm has a larger
variance and is less accurate in the com-Amazon graph. On the other
hand, the com-Amazon graph has a small value of the largest eigenvalue
which enables the BLC and the SRE algorithms to converge to the actual
value quickly.

As summarized in \cite{MahKri2006}, graphs which have high values of
the largest eigenvalues usually have a small diameter, good expansion
features and are more robust. Besides, the speed of propagation is
higher in graphs with a large spectral gap, the difference between the
first and second largest eigenvalues. Many social network graphs which
are of primary interest in BigData graph analytics have a small
diameter and good propagation properties; so, our algorithm is capable
of achieving a good performance on such graphs. 

\begin{figure*}[!th]
\centering 
       \includegraphics[width=2in]{legend}
       \\
\subfigure[{email-EuAll}]{
       \includegraphics[width=2.85in]{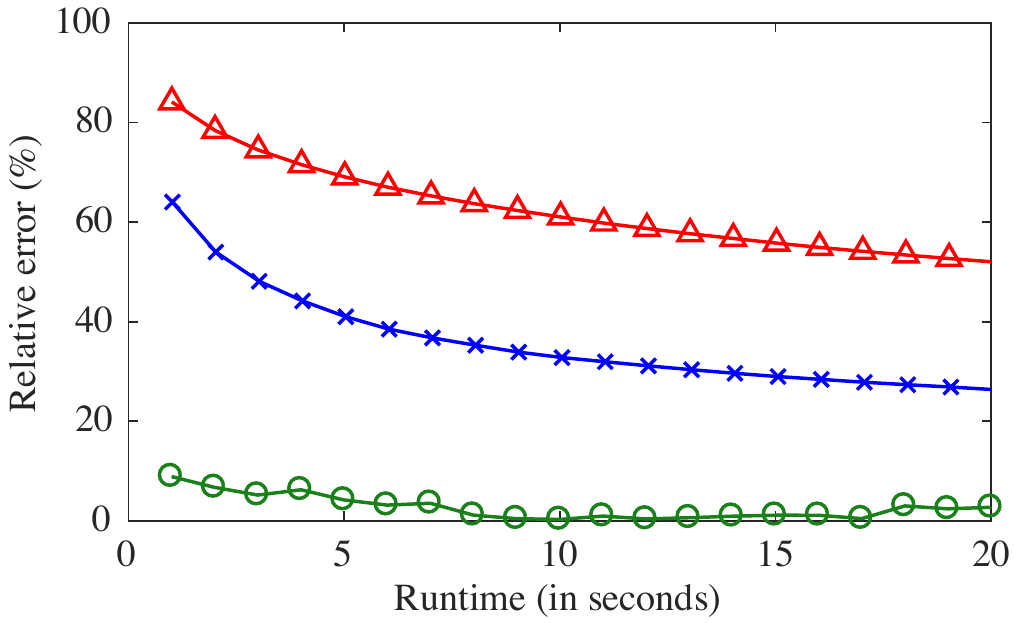}
       \label{fig:runtime1}
       }
       \subfigure[{com-Youtube}]{
       \includegraphics[width=2.85in]{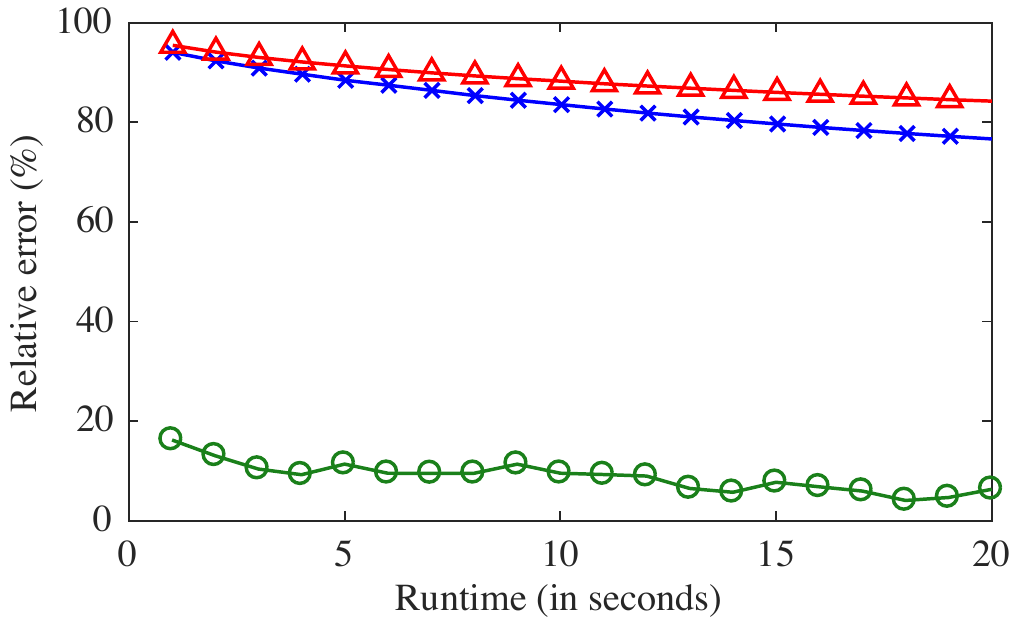}
       \label{fig:runtime2}
       }
              \subfigure[{loc-gowalla}]{
       \includegraphics[width=2.85in]{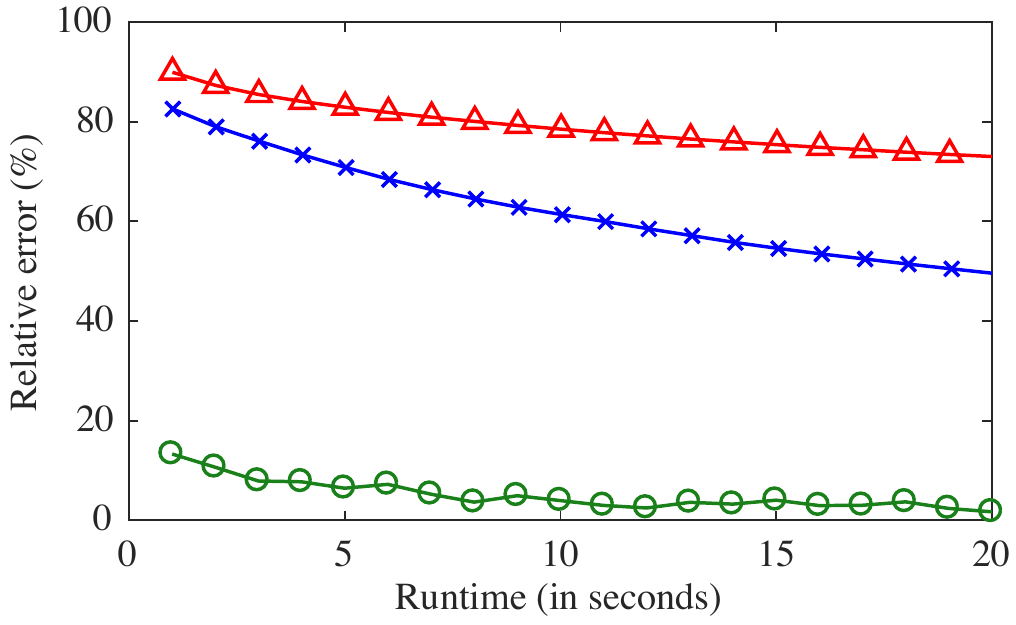}
       \label{fig:runtime4}
       }
       \subfigure[{com-Amazon}]{
       \includegraphics[width=2.85in]{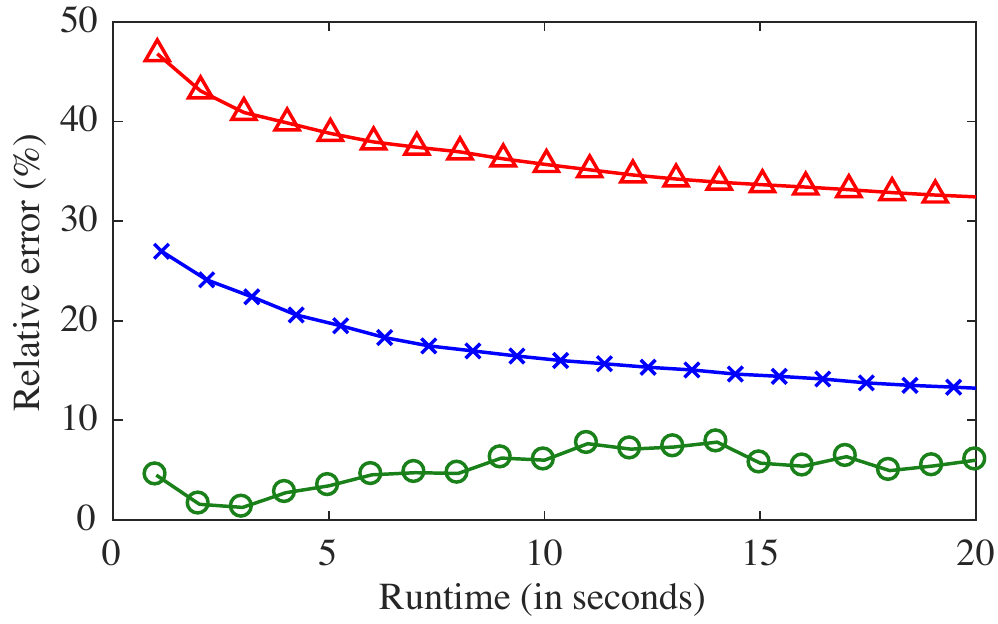}
       \label{fig:runtime3}
       }
\caption{Comparisons of the relative error in estimating the largest eigenvalue against the runtime in seconds.}
\label{fig:runtime}
\end{figure*}

\subsubsection{\textbf{Runtime}}
Both BLC and SRE try to sample the nodes with the largest eigenvalue
centrality in the graph. They calculate and update the score of each
neighboring node of the current sample set in order to select 
a node with the highest score. This leads to high computational
complexity. The cWalker-B algorithm, on the other hand, achieves a
significant improvement in the runtime by avoiding such score
calculations and using the less computationally intensive method of a
simple random walk.

We implemented the three algorithms, cWalker-B, BLC and SRE, in python
using $networkx$ and $igraph$ modules. All of the simulations were run
on an iMac with 16GB 1600MHz DDR3 memory and 2.7GHz Intel Core i5
processor. Here we assume that all of the graphs are stored on the
local machine, so the query time is neglected. 

Fig. \ref{fig:runtime} depicts the relative error of the estimate
reached by the three algorithms against the runtime, averaged over
100 independent runs. As shown in the figure, the cWalker-B algorithm
achieves much smaller relative errors for the same
runtime. Of particular interest is the case of the com-Amazon graph,
as plotted in Fig. \ref{fig:runtime3}, where it achieves better
accuracy within the same runtime. This is in contrast to the earlier
finding under the baseline of the number of queries (shown in
Fig. \ref{fig:amazon-error}), where our algorithm achieves a lower accuracy given the
same number of queries allowed. The cWalker-B algorithm is substantially
faster than the other two algorithms because, given the same amount of
runtime, it is able to process more information (visiting more nodes
in the random walk) and achieve better accuracy.  

\subsection{Results of estimating the two largest eigenvalues}
In this subsection, we show the results of estimating the two largest eigenvalues using the cWalker-C algorithm (as described in Algorithm  \ref{alg:cWalker-TopTwo}). For all of the experiments evaluating this algorithm, the accuracy target $\beta$ and the maximum length of closed walk $K$ were set to 0.01 and 30, respectively. 

Both SRE and BLC aim to collect a set of nodes which have the largest estimated eigenvalue centrality, and are not designed for estimating the second largest eigenvalue. However, based on interlacing results in spectral graph theory, where the eigenvalues of the full graph can be bounded using the eigenvalues of its subgraphs, the second largest eigenvalue of the sampled graph obtained by SRE and BLC can serve as a reference. 

\begin{figure*}[!t]
\centering
     
       \includegraphics[width=5in]{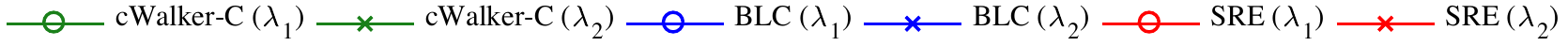}
       \\
    \subfigure[{email-EuAll}]{ 
       \includegraphics[width=2.85in]{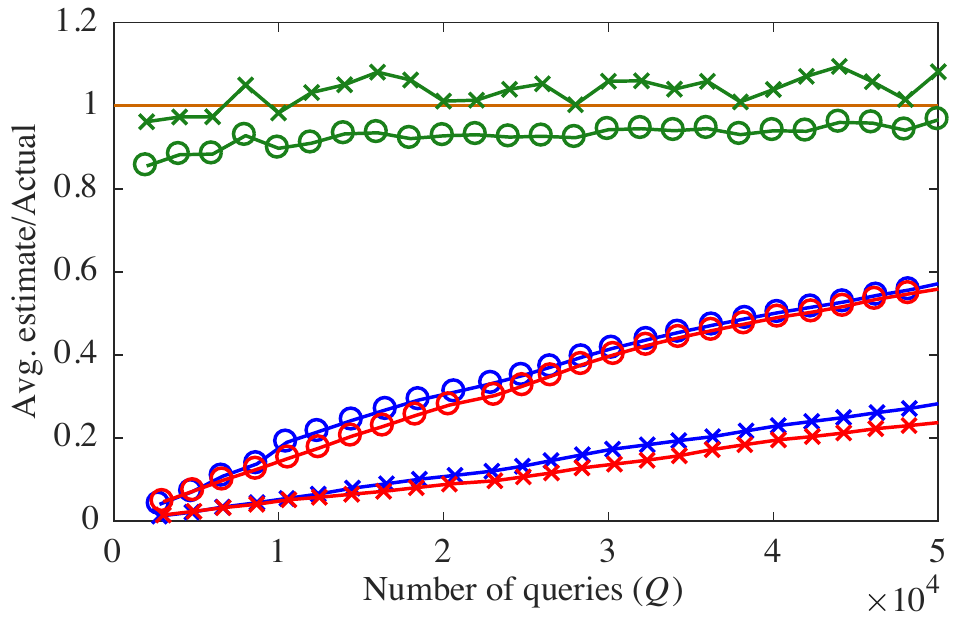} 
       }
      \subfigure[{com-Youtube}]{ 
       \includegraphics[width=2.85in]{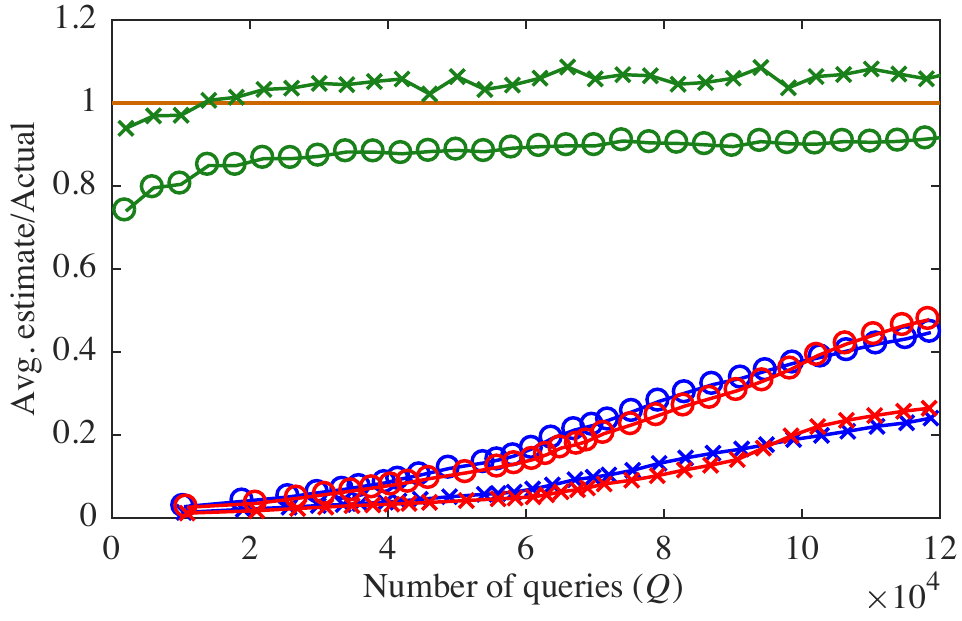}
       }
       \subfigure[{loc-gowalla}]{ 
       \includegraphics[width=2.85in]{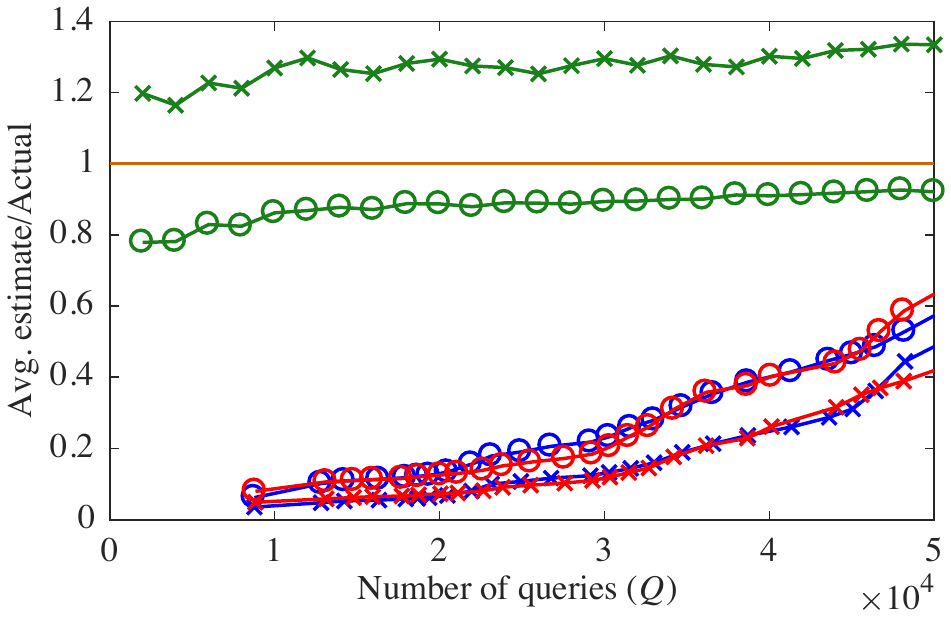}
       \label{fig:rel-gowalla-twoeig}
       }
              \subfigure[{com-Amazon}]{ 
      \includegraphics[width=2.85in]{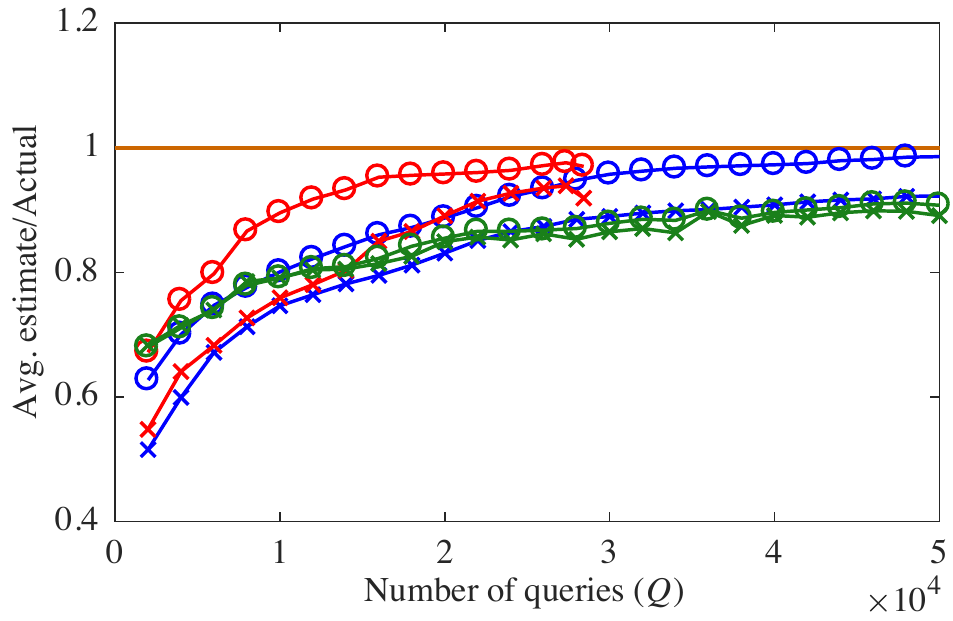}
      \label{fig:amazon-error-twoeig}
        }

    \caption{Comparison of the ratio of the average estimated value of
      $\lambda_1$ and $\lambda_2$ to their actual values. The brown line indicates 1.}\label{fig:accuracy-twoeig} 
\end{figure*}

Fig. \ref{fig:accuracy-twoeig} depicts the ratio of the average estimated $\lambda_1$ and $\lambda_2$ to their actual values for each of the four graphs with increasing number of queries. Similar to the results of the cWalker-B algorithm, cWalker-C also cannot outperform the other two algorithms in the com-Amazon graph. Since the ratio of $\lambda_2$ and $\lambda_1$ of the com-Amazon graph is very close to 1, the value of $k$ used in the estimation is large. Thus, in the com-Amazon graph, our algorithm takes more steps to converge. Except for the case of the com-Amazon graph, the cWalker-C algorithm achieves substantially better accuracy than BLC and SRE in the estimation of both $\lambda_1$ and $\lambda_2$. As plotted in the figure, in the cases of the email-EuAll, loc-gowalla and com-Youtube graphs, the estimates of $\lambda_2$ obtained by our algorithm do not show a clear convergence towards the actual values (e.g., plots do not converge to the brown line). Especially in Fig. \ref{fig:rel-gowalla-twoeig}, we can see that the accuracy in estimating $\lambda_2$ becomes lower with the increasing number of queries. The reason why the estimates of $\lambda_2$ obtained by our algorithm do not converge towards the ground truth is that the value of $k$ used for estimating $\lambda_2$ is not large enough. As shown in Eqn. (\ref{eqn:eigenTrace}), when $k$ is not large enough, $\lambda_1$ and $\lambda_2$ cannot become the dominate terms in the RHS of this equation. In other words, the value of $\sum_{i=3}^{|V|} {\lambda_i}^k$ is unnegligible and makes the approximation of $\lambda_2$ be overestimated. Similar to the case of $\lambda_1$, the reasonable value of $k$ for estimating $\lambda_2$ depends on the ratio of $\lambda_3$ and $\lambda_2$. If this ratio is close to 1, the value of $k$ has to be very large in order to get an accurate estimate of $\lambda_2$. However, in our algorithm, we use the approximation of $\lambda_1$ to estimate $\lambda_2$, so the value of $k$ used in the estimation of $\lambda_2$ must be no larger than the one used to obtain the estimate of $\lambda_1$. For a graph where the ratio of $\lambda_2$ and $\lambda_1$ is small, while the ratio of $\lambda_3$  and $\lambda_2$ is large (e.g., close to 1), the approximation of $\lambda_2$ obtained by our algorithm can be overestimated due to the use of a small value of $k$. To avoid this situation, we can set the minimum value of $k$ to a relatively large value.

\begin{figure*}[!th]
\centering 
       \includegraphics[width=5in]{legend-twoeig}
       \\
\subfigure[{email-EuAll}]{
       \includegraphics[width=2.85in]{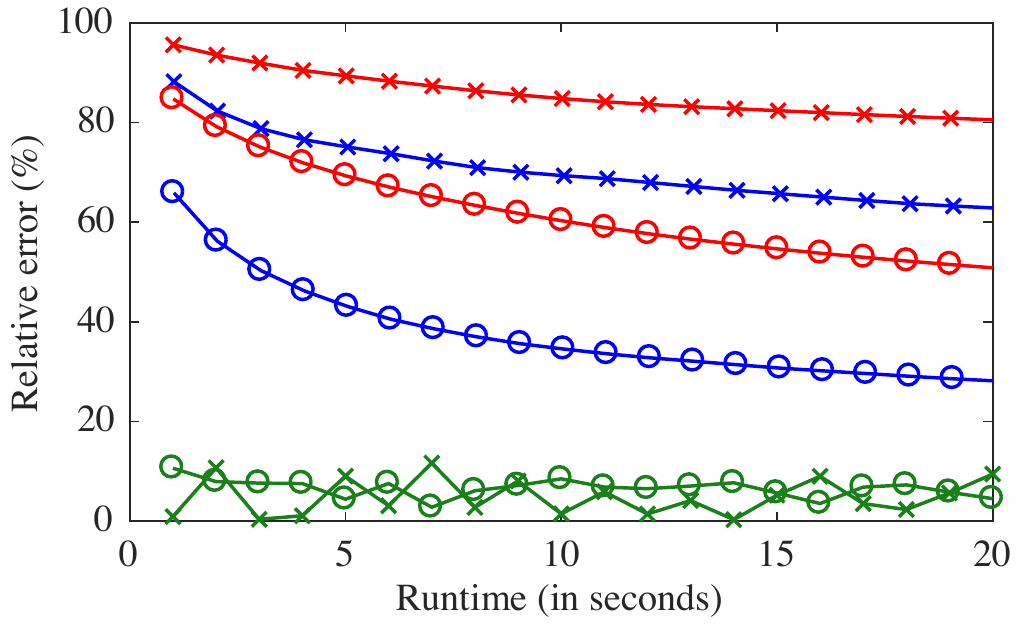}
       \label{fig:runtime1-two}
       }
       \subfigure[{com-Youtube}]{
       \includegraphics[width=2.85in]{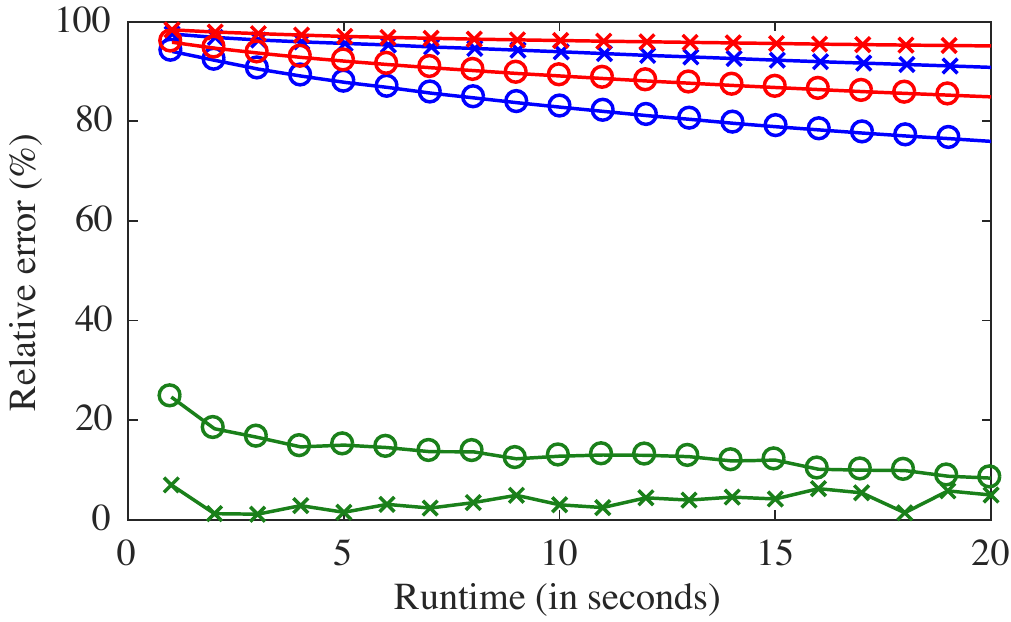}
       \label{fig:runtime2-two}
       }
              \subfigure[{loc-gowalla}]{
       \includegraphics[width=2.85in]{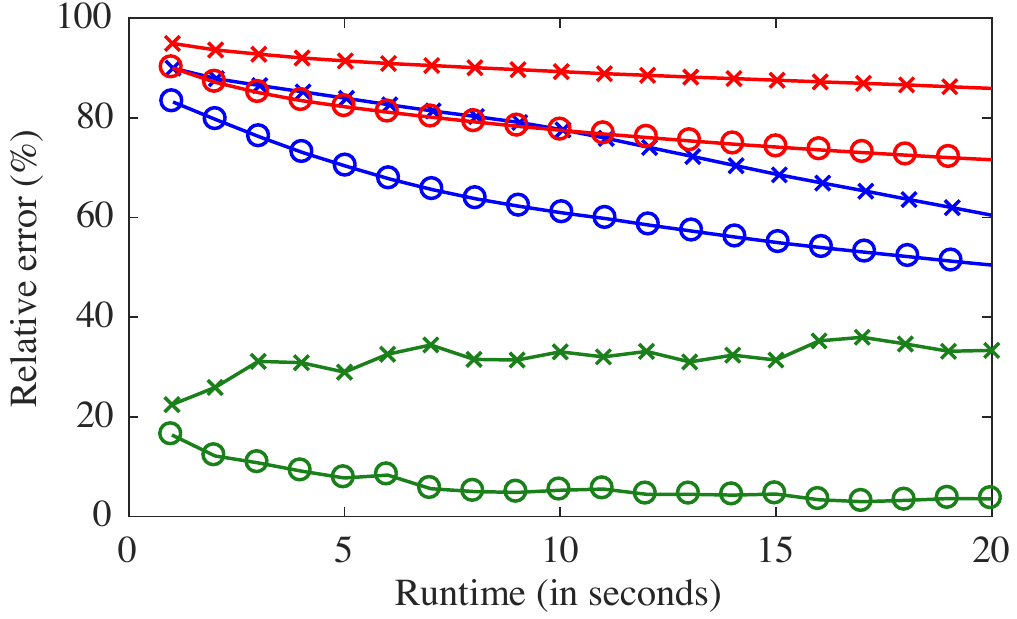}
       \label{fig:runtime4-two}
       }
       \subfigure[{com-Amazon}]{
       \includegraphics[width=2.85in]{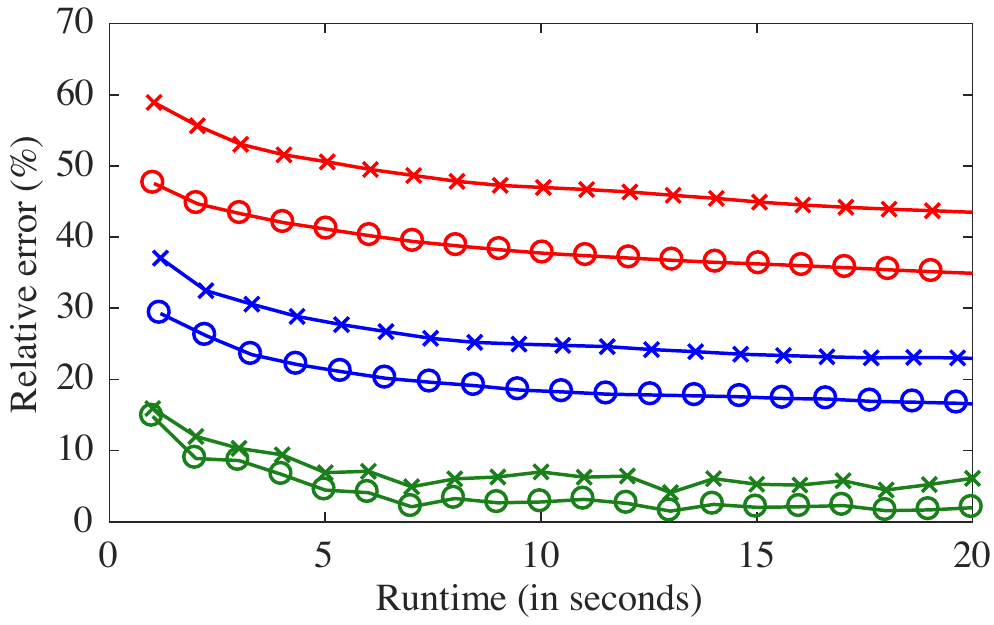}
       \label{fig:runtime3-twoeig}
       }
\caption{Comparison of the relative error in estimating the two largest eigenvalues against the runtime in seconds.}
\label{fig:runtime_twoeig}
\end{figure*}

Fig. \ref{fig:runtime_twoeig} plots the relative error of the estimates obtained by the three algorithms against the runtime, averaged over 100 independent runs. As plotted in the figure, the cWalker-C algorithm achieves much smaller relative errors under the same runtime. Similar to the results of the cWalker-B algorithm, the cWalker-C algorithm also achieves better accuracy within the same runtime in the com-Amazon graph. Combining with the results shown in Fig. \ref{fig:amazon-error-twoeig}, where our algorithm reaches a lower accuracy given the same number of queries allowed, we can say that the cWalker-C algorithm is faster than the other two algorithms. Moreover, as shown in Fig. \ref{fig:runtime4-two}, the relative error in estimating $\lambda_2$ obtained by our algorithm becomes larger with increasing amount of runtime. This is in line with the results shown in Fig. \ref{fig:rel-gowalla-twoeig}. The ratio of $\lambda_2$ and $\lambda_1$ of the loc-gowalla graph is 0.65, while its ratio of $\lambda_3$  and $\lambda_2$ is 0.95. The value of $k$ used for estimating $\lambda_2$ is not large enough, and thus leads to an overestimation of $\lambda_2$ in the case of the loc-gowalla graph. 

\subsection{Complexity}
\begin{table}[!t]
\centering
\caption{The complexity of the algorithms.}
\vspace{0.05in}
\label{tab:complexity}
\begin{tabular}{c|c|c} 
\Xhline{1pt}
  Algorithm&Time complexity&  Space complexity  \\ \hline
cWalker-B&$O( t \log \Delta)$& $O(K+\Delta)$ \\ \hline
cWalker-C&$O( t \Delta \log \Delta)$&$O(K\Delta)$ \\ \hline
SRE& $O(n\Delta(s+\Delta))$ &$O(s\Delta)$ \\ \hline
BLC& $O(\Delta^2 s^3)$& $O(s\Delta)$ \\ 
\Xhline{1pt}
\end{tabular}
\end{table}

Besides accuracy, our algorithms achieve a much better
performance in terms of the space complexity and the computational
costs. Table III summaries the complexity of the algorithms.

\subsubsection{\textbf{Computational complexity}}

Denote by $\Delta$ the maximum degree of a node in the graph. The complexity of checking the occurrence of a node in the 
neighborhood of another node is $O(\log \Delta)$, assuming the neighbors of each node are stored in a sorted list. In the cWalker-B
algorithm, at each step, we check if the node visited $k$ steps
earlier is in the neighborhood of the current node; this leads to the
complexity of $O(\log \Delta)$. For a random walk of length $t$, the complexity becomes $O(t\log \Delta)$. The cWalker-C algorithm needs to find common elements in the neighboring nodes of the first and the last nodes of a path; this takes $O(\Delta \log \Delta)$. So, its complexity is $O(t\Delta\log \Delta)$.


The runtime of SRE is highly influenced by the structure of the graph
and the selection of the starting node. Every time the sample graph is
updated, SRE updates the scores of all the corresponding
nodes, which can consume significant time for graphs with high average node degree. SRE is
a greedy algorithm, so the selection of the starting node is
especially determinative of the runtime. As presented in
\cite{ChuSet2015}, SRE takes $O(s\Delta+\Delta ^2)$ to select 
one node for removal from the sample subgraph and one node
for addition into it, where $s$ is the size of the sample graph. 
In the worst case, SRE may
have to visit the entire graph in order to make an estimate with sufficient accuracy which leads to a complexity of $O(n\Delta (s+ \Delta))$. 

BLC adopts a different metric to compute the score of the nodes, using the number of neighbors of a node in the sample graph (unlike SRE which uses the sum of the degrees of the node's neighbors.) 
The complexity of the BLC algorithm, as described in \cite{ChoGar1998}, is $O(\Delta^2 s^3)$. 

\subsubsection{\textbf{Space complexity}}
For the cWalker-B algorithm, as we visit nodes in the random walk, we keep
checking if one of the neighboring nodes of the current node is
identical to the previously visited node $k-1$ steps earlier. So we only
need to store the neighborhood information of one node and track up to $k$ steps of the random walk. The cWalker-B algorithm has a space complexity of
$O(K + \Delta)$, where $K$ is the maximum length of closed walk being checked in the algorithm. Note that $K$ is a small constant which is accepted as an input. 
As for the cWalker-C algorithm, we need to check the number of common neighbors between the first and the last node in paths of length $k-2$. To avoid querying the neighboring nodes of previously visited nodes again, the neighborhood information of previously visited $k-2$ nodes are stored. So the space complexity is $O(K \Delta)$

For the BLC and SRE algorithms,
however, at each step, both of them need to pick up a node which has
the highest score from the neighborhood of the current sample graph, so they have to store
both the sample graph and the score of each neighboring node of the
current sample graph which leads to a complexity of $O(s\Delta)$.

\section{Conclusion}
\label{sec:conclude}
In this paper, we present a series of new sampling algorithms 
which estimate the largest and the second largest eigenvalues of the
graph. Unlike previous methods which seek out nodes with high
eigenvalue centrality based on some score, our algorithm achieves a
significant improvement in computational efficiency by adopting an
entirely different approach. Our method is based on estimating the
number of closed walks of length $k$ by exploiting its relationship to
the $k$-th spectral moment of the graph.
Our results demonstrate that, on most graphs, our algorithms
achieve substantially better accuracy at a lower computational cost
than previously known algorithms.

Random walks on graphs had previously been used by graph sampling
algorithms to ascertain simple properties of graphs such as its
clustering co-efficient, motif statistics or centrality measures. This
paper offers hope that random walks can indeed be employed to sample
and estimate complex properties of graphs such as its eigenvalues. 

\ifCLASSOPTIONcompsoc
  \section*{Acknowledgments}
\else
  \section*{Acknowledgment}
\fi

This work was partially funded by the National Science Foundation
Award \#1250786.

\ifCLASSOPTIONcaptionsoff
  \newpage
\fi

\bibliographystyle{IEEEtran}
\bibliography{refs}

\end{document}